\documentclass[preprint,showpacs,preprintnumbers,amsmath,amssymb]{revtex4}

\usepackage{graphicx}% Include figure files
\usepackage{dcolumn}% Align table columns on decimal point
\usepackage{bm}% bold math
\usepackage{dsfont}
\usepackage{epsfig}
\usepackage{psfrag}
\begin{document}

\preprint{ }

\title{Electromagnetic Meson Form Factor\\ from a Relativistic Coupled-Channel Approach}
% Force line breaks with \\

\author{Elmar P. Biernat}
\email{elmar.biernat@uni-graz.at}
%\altaffiliation[Also at ]{Physics Department, XYZ University.}
%Lines break automatically or can be forced with \\

\author{Wolfgang Schweiger}%
\email{wolfgang.schweiger@uni-graz.at} \affiliation{ Institut
f\"ur Physik, Universit\"at Graz, A-8010 Graz, Austria}

\author{Kajetan Fuchsberger}
%\homepage{http://www.Second.institution.edu/~Charlie.Author}
\affiliation{ AB-OP-SPS Division, CERN, CH-1211 Geneve 23,
Switzerland } \email{kajetan.fuchsberger@cern.ch}

\author{William H. Klink}
%\homepage{http://www.Second.institution.edu/~Charlie.Author}
\affiliation{Department of Physics and Astronomy, University of
Iowa, Iowa City, IA, USA} \email{william-klink@uiowa.edu}

\date{\today}% It is always \today, today,
             %  but any date may be explicitly specified

\begin{abstract}
Point-form relativistic quantum mechanics is used to derive an
expression for the electromagnetic form factor of a pseudoscalar
meson for space-like momentum transfers. The elastic
scattering of an electron by a confined quark-antiquark pair is
treated as a relativistic two-channel problem for the $q\bar{q}e$
and $q\bar{q}e\gamma$ states. With the approximation that the total
velocity of the $q\bar{q}e$ system is conserved at (electromagnetic)
interaction vertices this simplifies to an eigenvalue problem for a
Bakamjian-Thomas type mass operator. After elimination of the
$q\bar{q}e\gamma$ channel the electromagnetic meson current and form
factor can be directly read off from the one-photon-exchange optical
potential. By choosing the invariant mass of the electron-meson
system large enough, cluster separability violations
become negligible.  An equivalence with the usual front-form
expression,  resulting from a spectator current in the $q^+=0$
reference frame, is established. The generalization of this
multichannel approach to electroweak form factors for an arbitrary
bound few-body system is quite obvious. By an appropriate extension of
the Hilbert space this approach is also able to accommodate
exchange-current effects.
\phantom{Point-form relativistic quantum mechanics is applied to
derive an expression for the electromagnetic form factor of a
pseudoscalar meson. To this aim the elastic scattering of an
electron by a confined quark-antiquark pair is treated as a
relativistic two-channel problem for the $q\bar{q}e$ and
$q\bar{q}e\gamma$ states. With the approximation that the total
velocity of the $q\bar{q}e$ system is conserved at
(electromagnetic) interaction vertices this simplifies to an
eigenvalue problem for a Bakamjian-Thomas type mass operator. }
\end{abstract}

% PACS, the Physics and Astronomy
                             % Classification Scheme.
%\keywords{Suggested keywords}%Use showkeys class option if keyword
                              %display desired

\maketitle

\section{\label{sec:introduction} Introduction}
Elastic electron-hadron scattering is an important source of
information on the internal structure of hadrons. Usually it is
treated in the one-photon-exchange approximation so that the
invariant scattering amplitude can be written as the contraction
of a (point-like) electron current with a hadron current times the
photon propagator~\footnote{More recently two-photon-exchange
effects have also been studied in connection with electron-nucleon
scattering~\cite{Blunden:2005ew,Kobushkin:2008ry}.}. For the
phenomenological analysis of electron-hadron scattering the most
general structure of the hadron current,  compatible with
Poincar\'e covariance and current conservation, is assumed. The
hadron current is a sum of independent 4-vectors multiplied by
Lorentz invariant functions, the hadron form factors. The form
factors are the observables which encode the electromagnetic
structure of the hadron. The kinematic quantities that are
available for constructing the hadron current are the 4-momenta of
the incoming and outgoing hadron. The form factors are thus
functions of the only Lorentz invariant variable one can build
from these 4-vectors, namely the 4-momentum-transfer squared.

The theoretical analysis of hadron form factors amounts to asking
how the electromagnetic current of the hadron may be expressed in
terms of the electromagnetic currents of the constituents. The
fact that the hadron current cannot be a simple sum of the
constituent currents has long been recognized
~\cite{Siegert:1937yt}. From the transformation properties of the
current operator $\hat{J}_\mu(x)$ under Poincar\'e transformations
it is quite obvious that the transformed operator
$\hat{J}^\prime_\mu(x^\prime)$ will, in general, be interaction
dependent as soon as some of the Poincar\'e generators contain
interaction terms. The binding forces must thus also show up in
the hadron current. How the binding interaction enters the hadron
current is further restricted by current conservation and the
physical condition that there be no renormalization of the hadron
charge. The latter means that the hadron charge should be equal to
the sum of the constituent charges, independent on whether the
binding forces are present or not. For relativistic quantum
mechanics of systems with a fixed number of particles the most
general form of Poincar\'e covariant tensor operators in a given
dynamical model has been derived in Ref.~\cite{Polyzou:1988tc}.
One of the conclusions of the authors is that relativity alone
does not provide any strong dynamical constraints on the
electromagnetic current operator. A whole class of current
operators for two- and three-particle bound states, which
satisfies the requirements of Poincar\'e covariance, current
conservation and cluster separability, has been formally
constructed in Ref.~\cite{lev:1995}, where the existence of
solutions for an arbitrary number of particles has also been
proved. For a comprehensive survey on the problem of constructing
the electromagnetic current operator of a bound few-body system
within the framework of relativistic quantum mechanics and the
different attempts to solve this problem we refer to the
introduction of Ref.~\cite{lev:1995}.

The procedure for constructing current operators proposed in
Ref.~\cite{lev:1995} is based on the point-form of relativistic
quantum mechanics. This form is characterized by the property that
all 4 generators of space-time translations are interaction
dependent, whereas the generators of Lorentz transformations stay
free of interactions. It  thus provides a natural starting point for the
construction of Lorentz-covariant tensor operators. Another issue
which is naturally addressed in the point form is the problem of cluster
separability, i.e. roughly speaking the property that arbitrary
subsystems of an interacting system should not interact with each
other, if they are separated by large space-like
distances~\cite{Sokolov:1977ym,Coester:1982vt,  Keister:1991sb}.
The cluster separability condition for the electromagnetic current
operator means that it must become the sum of subsystem current
operators if the interaction between the subsystems is turned off.
Cluster separability is closely related to the previously mentioned
requirement that the charge of the whole system should not be
renormalized by the interaction.

By making use of the equivalence of the different forms of
relativistic quantum mechanics the point-form results have also
been obtained in the instant and front forms. It should be noted
however, that all the results in Ref.~\cite{lev:1995} are rather
formal and no explicit application is presented. Subsequently the
whole form factor analysis has been redone in the front form of
relativistic quantum mechanics~\cite{Lev:1998qz} and applied to
the calculation of electromagnetic properties of the
deuteron~\cite{Lev:2000vm} and the pion~\cite{deMelo:2005cy}.

In the present paper we will also exploit the virtues of the point
form of relativistic quantum mechanics for the calculation of meson
form factors within constituent quark models. Our strategy, however,
differs from the one in Ref.~\cite{lev:1995}. Rather than starting with
the most general form of the electromagnetic current operator and
trying to satisfy the covariance, current-conservation and
cluster-separability constraints such that compatibility with a
particular interaction model is achieved, we  start with a
Poincar\'e invariant treatment of electron-hadron scattering in
which the dynamics of the exchanged photon is explicitly taken into
account. Since the one-photon-exchange optical potential is expected
to have the structure of a current-current interaction it should be possible
 to extract the hadron current. Once this is done it will be necessary to
investigate whether the resulting current has all the desired
properties and if not whether and how its deficiencies can be cured.

The general Poincar\'e invariant framework which we use to
describe electron-hadron scattering will be presented in
Sec.~\ref{sec:massop}. It is a relativistic two-channel formalism
for a Bakamjian-Thomas type mass operator~\cite{Bakamjian:1953kh,
Keister:1991sb} in a velocity-state
representation~\cite{Klink:1998zz}. The two channels are either
the electron-meson and electron-meson photon channels or the
electron-quark-antiquark and electron-quark-antiquark-photon
channels, depending on whether the problem is considered on the
hadronic or constituent level. The field theoretical vertices for
the emission and absorption of a photon are implemented in the
Bakamjian-Thomas type framework as suggested in
Ref.~\cite{Klink:2000pp}.
%
% and have already been applied to the
% calculation of the vector-meson spectrum within the chiral
%constituent-quark model~\cite{Krassnigg:2003gh}.
%
In Sec.~\ref{subsec:mesonmass} we will first demonstrate on the
hadronic level that the one-photon-exchange optical potential has
indeed the structure of a current-current interaction with a
conserved meson current that transforms like a 4-vector under
Lorentz transformations. This holds even if one allows for a
phenomenological vertex form factor at the photon-meson vertex.
The whole calculation will be repeated on the constituent level
with an instantaneous confining potential between quark and
antiquark in Sec.~\ref{subsec:clustermass}. In this way a
microscopic meson current which is conserved and behaves like a
4-vector can again be extracted from the one-photon-exchange
optical potential. By comparing appropriate matrix elements of the
one-photon-exchange optical potentials on the hadronic and on the
constituent level the electromagnetic form factor will be
identified in Sec.~\ref{sec:formfac}. But it will turn out not to
satisfy the
 required cluster properties.   By going to the infinite
momentum frame of the meson the form factor  becomes a simple
analytical expression and satisfies the cluster separability property.
As a first numerical application the electromagnetic pion form factor
will be calculated with a simple harmonic-oscillator wave
function. In Sec.~\ref{subsec:frontform} it will be shown that our
form factor result is equivalent to the usual front-form
expression, resulting from a spectator current in the $q^+=0$
frame. Section~\ref{subsec:pfspectator} is devoted to a comparison
with the so called \lq\lq point-form spectator
model\rq\rq~\cite{Wagenbrunn:2000es,Boffi:2001zb} -- which is another
procedure for calculating electromagnetic hadron form factors in
point-form quantum mechanics. A summary and the possibility for
further generalizations and applications of the formalism
presented in this paper will be given in
Sec.~\ref{sec:conclusion}.

\section{\label{sec:massop}Mass operator for electron-meson scattering}
As already mentioned in the introduction, we will treat the
electromagnetic scattering of an electron by a pseudoscalar meson as
a two-channel problem for an appropriately defined mass operator.
This mass operator acts on a Hilbert space which is the direct sum of the
incoming and outgoing electron-meson states and the intermediate
electron-meson-photon states. Decomposing an arbitrary mass
eigenstate of the system $\vert \psi \rangle$ into components
belonging to these two channels, $\vert \psi \rangle = \vert \psi_{e
M} \rangle + \vert \psi_{e M \gamma} \rangle$, the eigenvalue
equation for the mass operator $\hat{M} \vert \psi \rangle = m \vert
\psi \rangle$ becomes a system of coupled equations for the
respective components:
\begin{eqnarray}\label{eq:coupchann}
\hat{M}_{e M}\vert \psi_{e M} \rangle + \hat{K}^\dag \vert \psi_{e M
\gamma}
\rangle & = & m \vert \psi_{e M} \rangle\, ,\nonumber \\
\hat{K} \vert \psi_{e M} \rangle + \hat{M}_{e M \gamma} \vert
\psi_{e M \gamma} \rangle & = & m \vert \psi_{e M \gamma}\rangle\,
.
\end{eqnarray}
$\hat{K}$ and $\hat{K}^\dag$ are vertex operators that are
responsible for the emission and absorption of a photon,
$\hat{M}_{e M}$ and $\hat{M}_{e M \gamma}$ are mass operators for
the free electron-meson and electron-meson-gamma systems,
respectively. We can solve the second equation for $\vert \psi_{e
M \gamma}\rangle$ and insert the resulting expression into the
first one to end up with a non-linear eigenvalue equation for
$\vert \psi_{e M}\rangle$:
\begin{equation}
\label{eq:DynamicalEquationM} \left(\hat{M}_{e M }-m\right)\vert
\psi_{e M} \rangle = \hat{K}^\dag { \left(\hat{M}_{e M \gamma}
-m\right) }^{-1} \hat{K} \vert \psi_{e M} \rangle =: \hat{
V}_\mathrm{opt}(m)\, \vert \psi_{e M} \rangle\, .
\end{equation}
The right-hand side of this equation describes the action of the
one-photon-exchange  optical potential $\hat{V}_\mathrm{opt}$ on
the $\vert \psi_{e M}\rangle$ state.

Particular matrix elements of the optical potential are
needed for the identification and extraction of the electromagnetic
meson form factor. We derive them first for the case where the
composite nature of the meson is taken into account by means of a
phenomenological form factor at the photon-meson vertex. In the
sequel we will redo the calculation for the case where the meson is
treated as a quark-antiquark bound state that has the quantum
numbers of the meson. In the latter case the photon couples directly
to the point-like constituent (anti)quark. A comparison of both
cases will finally allow us to identify the electromagnetic current
of the meson and to express the electromagnetic meson form factor in
terms of the quark-antiquark bound-state wave function.

\subsection{\label{subsec:mesonmass}Hadronic level}
Relativistic invariance of our quantum mechanical description of
electron-meson scattering is guaranteed if one is able to find a
realization of the Poincar\'e algebra in terms of operators that
act on the direct sum of the $eM$ and $eM\gamma$ Hilbert spaces.
This can be achieved by means of the, so called, \lq\lq
Bakamjian-Thomas construction\rq\rq~\cite{Bakamjian:1953kh}. Its
point-form version amounts to the assumption that the free
4-velocity operator $\hat{V}^\mu_\mathrm{free}$ can be factored
out of the (interacting) 4-momentum operator,
\begin{equation}\label{eq:BTP}
\hat{P}^\mu = \hat{P}^\mu_\mathrm{free} +\hat{P}^\mu_\mathrm{int} =
(\hat{M}_\mathrm{free}+\hat{M}_\mathrm{int})\,
\hat{V}^\mu_\mathrm{free}\, ,
\end{equation}
so that one can concentrate on studying the mass operator
$\hat{M}=\hat{M}_\mathrm{free}+\hat{M}_\mathrm{int}$. The
Poincar\'e algebra is satisfied provided that the interacting part
of the mass operator $\hat{M}_\mathrm{int}$ is a Lorentz scalar
and commutes with $\hat{V}^\mu_\mathrm{free}$. For this kind of
construction it is advantageous to represent the mass operator and
the Poincar\'e generators in a basis that consists of velocity
states~\cite{Klink:1998zz}. An $n$-particle velocity state $\vert
v; \vec{k}_1, \mu_1; \vec{k}_2, \mu_2;\dots; \vec{k}_n, \mu_n
\rangle $ is simply obtained by starting from a multiparticle
momentum state in its rest frame and boosting it to overall
4-velocity $v$ ($v_\mu v^\mu = 1$) by means of a canonical spin
boost $B_c(v)$~\cite{Keister:1991sb}, i.e.
\begin{equation}\label{eq:velstat}
\vert v; \vec{k}_1, \mu_1; \vec{k}_2, \mu_2;\dots; \vec{k}_n, \mu_n
\rangle = \hat{U}_{B_c(v)} \, \vert \vec{k}_1, \mu_1; \vec{k}_2,
\mu_2;\dots; \vec{k}_n, \mu_n \rangle \quad \mathrm{with} \quad
\sum_{i=1}^n \vec{k}_i=0\, .
\end{equation}
The $\mu_i$s denote the spin projections of the individual
particles. By construction one of the $\vec{k}_i$s is redundant.
In the following we will make extensive use of orthogonality and
completeness relations of the velocity states. For a system of
$n\geq 2$ particles with 4-momenta $k_i$, masses $m_i$, energies
$\omega_{k_i}:= \sqrt{m_i^2+\vec{k}_i^2}$, and spin projections
$\mu_i$, $i=1,\dots,n$, we have the orthogonality relation
\begin{eqnarray}\label{eq:vnorm}
\lefteqn{\langle v^\prime; \vec{k}_1^\prime, \mu_1^\prime;
\vec{k}_2^\prime, \mu_2^\prime;\dots; \vec{k}_n^\prime, \mu_n^\prime
\vert \, v; \vec{k}_1, \mu_1; \vec{k}_2, \mu_2;\dots; \vec{k}_n,
\mu_n \rangle  \nonumber}\\ & & = v_0 \,
\delta^3(\vec{v}^\prime-\vec{v})\, \frac{(2 \pi)^3 2
\omega_{k_n}}{\left( \sum_{i=1}^n \omega_{k_i}\right)^3} \left(
\prod_{i=1}^{n-1} (2\pi)^3 2 \omega_{k_i}
\delta^3(\vec{k}_i^\prime-\vec{k}_i)\right) \left( \prod_{i=1}^{n}
\delta_{\mu_i^\prime \mu_i}\right)\, .
\end{eqnarray}
Here we have taken (without loss of generality) the $n$th momentum
to be redundant. The corresponding completeness relation reads:
\begin{eqnarray}\label{eq:vcompl}
\mathds{1}_{1,2,\dots,n}&=&\sum_{\mu_1=-j_1}^{j_1}
\sum_{\mu_2=-j_2}^{j_2} \dots \sum_{\mu_n=-j_n}^{j_n} \int
\frac{d^3 v}{(2\pi)^3 v_0} \left(
\prod_{i=1}^{n-1}\frac{d^3k_i}{(2 \pi)^3 2 \omega_{k_i}}
\right)\frac{\left(\sum_{i=1}^n \omega_{k_i}\right)^3}{2
\omega_{k_n}}\nonumber\\ & & \times \vert v; \vec{k}_1, \mu_1;
\vec{k}_2, \mu_2;\dots; \vec{k}_n, \mu_n \rangle \langle  v;
\vec{k}_1, \mu_1; \vec{k}_2, \mu_2;\dots; \vec{k}_n, \mu_n\vert\,
,
\end{eqnarray}
where $j_i$ is the spin of the $i$th particle. One of the big
advantages of velocity states as compared with the usual momentum states
is that under a Lorentz transformation $\Lambda$ the  Wigner rotation is
the same for all particles, i.e.
\begin{eqnarray}\label{eq:vstateboost}
\hat{U}_\Lambda \vert v; \vec{k}_1, \mu_1; \vec{k}_2, \mu_2;\dots;
\vec{k}_n, \mu_n \rangle = \sum_{\mu_1^\prime,
\mu_2^\prime,\dots,\mu_n^\prime}\, \left( \prod_{i=1}^n \,
D^{j_i}_{\mu_i^\prime \mu_i}(R_{\mathrm W}(v,\Lambda)) \right)
\nonumber\\ \times \vert \Lambda v; \overrightarrow{R_{\mathrm
W}(v,\Lambda)k}_1, \mu_1^\prime; \overrightarrow{R_{\mathrm
W}(v,\Lambda)k}_2, \mu_2^\prime;\dots; \overrightarrow{R_{\mathrm
W}(v,\Lambda)k}_n, \mu_n^\prime \rangle\, ,
\end{eqnarray}
with the Wigner-rotation matrix
\begin{equation}\label{eq:wignerrot}
R_{\mathrm W}(v,\Lambda) = B_c^{-1}(\Lambda v)\Lambda B_c(v)\, .
\end{equation}

In a velocity-state basis the Bakamjian-Thomas type 4-momentum
operator, Eq.~(\ref{eq:BTP}), becomes diagonal in the 4-velocity
$v$. This is a special feature of the Bakamjian-Thomas construction
which, in general, does not hold for arbitrary interacting
relativistic quantum theories. It is, in particular, not possible to
factorize the 4-momentum operator of an interacting point-form
quantum field theory as in
Eq.~(\ref{eq:BTP})~\cite{Biernat:2007sz}. The vertex operators
$\hat{K}$ and $\hat{K}^\dag$ in Eq.~(\ref{eq:DynamicalEquationM}),
which are responsible for photon emission and absorption, can
therefore not directly be taken from point-form quantum
electrodynamics.  One rather has to make the approximation that the
total 4-velocity of the system is conserved at the electromagnetic
vertices to end up with a Bakamjian-Thomas type mass operator. In
Ref.~\cite{Klink:2000pp} it has been demonstrated in some detail
that this is a way to implement general field theoretical vertex
interactions into a Bakamjian-Thomas type framework. If we denote
the velocity states of the $e M$ and $e M \gamma$ systems by $\vert
v; \vec{k}_e, \mu_e; \vec{k}_M \rangle$ and $\vert v; \vec{k}_e,
\mu_e; \vec{k}_M; \vec{k}_{\gamma}, \mu_{\gamma} \rangle$,
respectively, the (velocity conserving) electromagnetic vertex
interaction takes on the form~\cite{Klink:2000pp}
\begin{eqnarray}\label{eq:vertex1}
\langle v^\prime; \vec{k}_e^\prime, \mu_e^\prime;
\vec{k}_M^\prime; \vec{k}_{\gamma}^\prime, \mu_{\gamma}^\prime
\vert\!\! &\hat{K}& \!\!\vert v; \vec{k}_e, \mu_e; \vec{k}_M
\rangle = \langle v; \vec{k}_e, \mu_e; \vec{k}_M \vert
\,\hat{K}^\dag\, \vert v^\prime; \vec{k}_e^\prime, \mu_e^\prime;
\vec{k}_M^\prime; \vec{k}_{\gamma}^\prime, \mu_{\gamma}^\prime
\rangle^\ast
\nonumber\\
&=& v_0 \, \delta^3(\vec{v}^\prime-\vec{v})\,
\frac{(2\pi)^3}{\sqrt{(\omega_{k_e^\prime}+\omega_{k_M^\prime}+
\omega_{k_{\gamma}^\prime})^3}
\sqrt{(\omega_{k_e}+\omega_{k_M^{\phantom{\prime}}})^3}}
\\ & &\times  \langle v^\prime; \vec{k}_e^\prime, \mu_e^\prime;
\vec{k}_M^\prime; \vec{k}_{\gamma}^\prime, \mu_{\gamma}^\prime
\vert\left( f(\Delta m)\, \hat{\mathcal{L}}_{\mathrm{int}}^{M
\gamma}(0) + \hat{\mathcal{L}}_{\mathrm{int}}^{e \gamma}(0)
\right) \vert v; \vec{k}_e, \mu_e; \vec{k}_M \rangle\, , \nonumber
\end{eqnarray}
with $\hat{\mathcal{L}}_{\mathrm{int}}^{M \gamma}(x)$ and
$\hat{\mathcal{L}}_{\mathrm{int}}^{e \gamma}(x)$ representing the
usual interaction densities for scalar and spinor quantum
electrodynamics~\cite{Bjorken:1964}. This is a very natural way to
introduce field theoretical vertex interactions into the
Bakamjian-Thomas framework, but as a drawback the mass operator
does not cluster properly. But as will be shown, if the invariant
mass of the electron-meson system is made sufficiently large, the
effect of the wrong cluster separability property is eliminated.
The vertex form factor $f(\Delta m)$ depends on the magnitude of
the difference of the invariant masses in the initial and final
states, $\Delta m = \vert\, \omega_{k_e}+\omega_{k_M}-
\omega_{k_e^\prime}-\omega_{k_M^\prime}-\omega_{k_{\gamma}^\prime}
\vert$. It is introduced at this place to account for the
composite nature of the meson. Moreover, it  also serves to partly
compensate for the neglect of the off-diagonal terms in the
4-velocity and to regulate the integrals if necessary. But since
this is not our primary goal, we do not  introduce a second form
factor at the electron-photon vertex. After evaluation of the
matrix elements on the right-hand side of Eq.~(\ref{eq:vertex1})
the matrix elements of the vertex operator
become~\cite{Fuchsberger:2007}
\begin{eqnarray}\label{eq:vertex}\lefteqn{
\langle v^\prime; \vec{k}_e^\prime, \mu_e^\prime; \vec{k}_M^\prime;
\vec{k}_{\gamma}^\prime, \mu_{\gamma}^\prime \vert \, \hat{K}  \vert
v; \vec{k}_e, \mu_e; \vec{k}_M \rangle \nonumber }\\ &=& v_0 \,
\delta^3(\vec{v}^\prime-\vec{v})\,
\frac{(2\pi)^3}{\sqrt{(\omega_{k_e^\prime}+\omega_{k_M^\prime}+
\omega_{k_{\gamma}^\prime})^3}
\sqrt{(\omega_{k_e}+\omega_{k_M^{\phantom{\prime}}})^3}}\, (-1)\nonumber\\
&&\times \left[ \, Q_e\,
\bar{u}_{\mu_e^\prime}(\vec{k}_e^\prime)\gamma_\nu
u_{\mu_e}(\vec{k}_e)\,
\epsilon^\nu(\vec{k}_{\gamma}^\prime,\mu_{\gamma}^\prime)\, (2
\pi)^3
2 \omega_{k_M} \delta^3(\vec{k}_M^\prime - \vec{k}_M)\right. \\
& & \left. \phantom{\times} + Q_M \, f(\Delta m)\, (k_M^\prime +
k_M)_\nu \,
\epsilon^\nu(\vec{k}_{\gamma}^\prime,\mu_{\gamma}^\prime)\, (2
\pi)^3 2 \omega_{k_e} \delta^3(\vec{k}_e^\prime - \vec{k}_e)
\right]\, , \nonumber
\end{eqnarray}
where $\epsilon(\vec{k}_{\gamma},\mu_{\gamma})$ represent
appropriately orthonormalized photon polarization vectors which
satisfy the completeness relation
\begin{equation}\label{eq:polcomp}
\sum_{\mu_{\gamma}=0}^3
\epsilon_{\mu}(\vec{k}_{\gamma},\mu_{\gamma})(-g^{\mu_\gamma
\mu_\gamma})\, \epsilon_{\nu}^\ast (\vec{k}_{\gamma},\mu_{\gamma})
= - g_{\mu \nu}\, .
\end{equation}
Note that due to this definition of photon polarization states the
orthogonality and completeness relations, Eqs.~(\ref{eq:vnorm}) and
(\ref{eq:vcompl}), have to be modified for each occurring photon
such that $\delta_{\mu_\gamma \mu_\gamma^\prime} \rightarrow
(-g^{\mu_\gamma \mu_\gamma^\prime})$  and $\sum_{\mu_\gamma}
\rightarrow \sum_{\mu_\gamma} (-g^{\mu_\gamma \mu_\gamma})$.

What we want to calculate are matrix elements of the optical
potential $\hat{ V}_\mathrm{opt}$ between velocity states of the $e
M$ system. To this aim we insert the velocity-state completeness
relation for the $e M \gamma$ system (cf. Eq.~(\ref{eq:vcompl}))
into Eq.~(\ref{eq:DynamicalEquationM}) and take matrix elements
between $e M$ velocity states
\begin{eqnarray}\label{eq:voptmatr1}
\lefteqn{\langle v^\prime; \vec{k}_e^\prime, \mu_e^\prime;
\vec{k}_M^\prime \vert\, \hat{ V}_\mathrm{opt}(m)\, \vert v;
\vec{k}_e, \mu_e; \vec{k}_M \rangle } \nonumber \\ &=& \langle
v^\prime; \vec{k}_e^\prime, \mu_e^\prime; \vec{k}_M^\prime \vert \,
\hat{K}^\dag\, \left(\hat{M}_{e M \gamma} -m\right)^{-1}
\mathds{1}_{eM\gamma}^{\prime \prime}\, \hat{K} \vert v;
\vec{k}_e, \mu_e; \vec{k}_M \rangle \\
&=& \sum_{\mu_e^{\prime\prime},\, \mu_\gamma^{\prime\prime}} \int
\frac{d^3 v^{\prime\prime}}{(2 \pi)^3 v_0^{\prime\prime}} \int
\frac{d^3 k_e^{\prime\prime}}{(2 \pi)^3 2
\omega_{k_e^{\prime\prime}}} \int \frac{d^3 k_M^{\prime\prime}}{(2
\pi)^3 2 \omega_{k_M^{\prime\prime}}}
\frac{(\omega_{k_e^{\prime\prime}}+\omega_{k_M^{\prime\prime}}+
\omega_{k_\gamma^{\prime\prime}})^3}{2
\omega_{k_\gamma^{\prime\prime}}} \nonumber\\ & & \times \langle
v^\prime; \vec{k}_e^\prime, \mu_e^\prime; \vec{k}_M^\prime \vert
\, \hat{K}^\dag\,\vert v^{\prime \prime} ; \vec{k}_e^{\prime
\prime} , \mu_e^{\prime \prime} ; \vec{k}_M^{\prime \prime} ;
\vec{k}_{\gamma}^{\prime \prime} , \mu_{\gamma}^{\prime \prime}
\rangle \nonumber\\ & & \times \left(\omega_{k_e^{\prime \prime}}+
\omega_{k_M^{\prime \prime}}+ \omega_{k_\gamma^{\prime \prime}}
-m\right)^{-1} \, \langle v^{\prime \prime} ; \vec{k}_e^{\prime
\prime} , \mu_e^{\prime \prime} ; \vec{k}_M^{\prime \prime} ;
\vec{k}_{\gamma}^{\prime \prime} , \mu_{\gamma}^{\prime \prime}
\vert \, \hat{K} \vert v; \vec{k}_e, \mu_e; \vec{k}_M \rangle \, .
\nonumber
\end{eqnarray}
Here we have exploited that $\vert v^{\prime \prime} ;
\vec{k}_e^{\prime \prime} , \mu_e^{\prime \prime} ;
\vec{k}_M^{\prime \prime} ; \vec{k}_{\gamma}^{\prime \prime} ,
\mu_{\gamma}^{\prime \prime} \rangle$ are eigenstates of the free
$e M \gamma$ mass operator
\begin{equation}
\hat{M}_{e M \gamma} \vert\, v^{\prime \prime} ; \vec{k}_e^{\prime
\prime} , \mu_e^{\prime \prime} ; \vec{k}_M^{\prime \prime} ;
\vec{k}_{\gamma}^{\prime \prime} , \mu_{\gamma}^{\prime \prime}
\rangle = (\omega_{k_e^{\prime \prime}}+ \omega_{k_M^{\prime
\prime}}+ \omega_{k_\gamma^{\prime \prime}} ) \vert\,  v^{\prime
\prime} ; \vec{k}_e^{\prime \prime} , \mu_e^{\prime \prime} ;
\vec{k}_M^{\prime \prime} ; \vec{k}_{\gamma}^{\prime \prime} ,
\mu_{\gamma}^{\prime \prime} \rangle \, .
\end{equation}
If one now makes use of Eq.~(\ref{eq:vertex}), a corresponding
relation for $\hat{K}^\dag$, and the completeness relation,
Eq.~(\ref{eq:polcomp}), for the photon polarization vectors, the
integrals in Eq.~(\ref{eq:voptmatr1}) can be done by means of the
delta functions and one ends up with
\begin{eqnarray}\label{eq:voptmatr2}
\lefteqn{\langle v^\prime; \vec{k}_e^\prime, \mu_e^\prime;
\vec{k}_M^\prime \vert\, \hat{ V}_\mathrm{opt}(m)\, \vert v;
\vec{k}_e, \mu_e; \vec{k}_M \rangle } \nonumber \\ &=& v_0 \delta^3
(\vec{v}^{\, \prime} - \vec{v}\, )\frac{(2 \pi)^3
}{\sqrt{(\omega_{k_e^{\prime}}+\omega_{k_M^{\prime}})^3}
\sqrt{(\omega_{k_e^{\phantom{\prime}}}+\omega_{k_M^{\phantom{\prime}}})^3}}\nonumber\\
& & \times Q_e\, \bar{u}_{\mu_e^\prime}(\vec{k}_e^\prime)\gamma_\mu
u_{\mu_e}(\vec{k}_e)\,  \frac{(-g^{\mu \nu})}{2
\omega_{k_\gamma^{\prime\prime} }} \, Q_M (k_M^\prime + k_M)_\nu \nonumber\\
& & \times \left[ \frac{f(\Delta
m)}{\omega_{k_M^\prime}+\omega_{k_e}+\omega_{k_\gamma^{\prime\prime}}-m}+\frac{f(\Delta
m^\prime)}{\omega_{k_M}+\omega_{k_e^\prime}+\omega_{k_\gamma^{\prime\prime}}-m}
\right]\, .
\end{eqnarray}
Here $\Delta m = \vert\, \omega_{k_M}+\omega_{k_e}
-\omega_{k_M^\prime}-\omega_{k_e}-\omega_{k_\gamma^{\prime\prime}}\vert$
and $\Delta m^\prime = \vert\,
\omega_{k_M}+\omega_{k_e^\prime}+\omega_{k_\gamma^{\prime\prime}}
-\omega_{k_M^\prime}-\omega_{k_e^\prime}\vert$, respectively. The
two terms between square brackets correspond to the two possible
time orderings, i.e. photon emission by the meson with subsequent
absorption by the electron and vice versa. In evaluating the matrix
elements of $\hat{V}_\mathrm{opt}$ we have neglected self-energy
contributions in which the photon is emitted and absorbed by the
same particle. Although all double-primed variables should be
eliminated by now, we have kept $\omega_{k_\gamma^{\prime\prime}}$
for better readability. Since $\vec{k}_\gamma^{\prime\prime}=\pm
(\vec{k}_e^\prime - \vec{k}_e)$ differs only in the sign for the two
time orderings, $\omega_{k_\gamma^{\prime\prime}}=
\vert\,(\vec{k}_e^\prime - \vec{k}_e) \vert$ is independent of the
time ordering.

Eq.~(\ref{eq:voptmatr2}) represents a general matrix element of
the one-photon-exchange optical potential. Electromagnetic hadron
form factors, however, are usually extracted from the elastic
electron-hadron scattering amplitude, calculated in the
one-photon-exchange approximation. This means that we can restrict
our considerations to on-shell matrix elements of
$\hat{V}_\mathrm{opt}$ for which
$m=\omega_{k_M}+\omega_{k_e}=\omega_{k_M^\prime}+\omega_{k_e^\prime}$
and $\omega_{k_M}=\omega_{k_M^\prime}$,
$\omega_{k_e}=\omega_{k_e^\prime}$~\footnote{Note that the
velocity-state representation is associated with center-of-mass
kinematics, c.f. Eq.~(\ref{eq:velstat}).}. As a consequence we
have $\Delta m = \Delta m^\prime =
\omega_{k_{\gamma}^{\prime\prime}} = \vert \vec{k}_M^\prime -
\vec{k}_M \vert = \sqrt{-q_\mu q^\mu}=Q$, with $q^\mu = k_M^\prime
- k_M$ being the 4-momentum transfer between incoming and outgoing
meson. With these relations the two terms in square brackets can
be combined and the on-shell matrix elements of the optical
potential can be expressed as a contraction of the electromagnetic
electron current
\begin{equation}\label{eq:elcurr}
j_\mu (\vec{k}_e^\prime, \mu_e^\prime;\vec{k}_e, \mu_e) = Q_e\,
\bar{u}_{\mu_e^\prime}(\vec{k}_e^\prime)\gamma_\mu
u_{\mu_e}(\vec{k}_e)
\end{equation}
with an electromagnetic meson current
\begin{equation}\label{eq:mcurrphen}
J_\nu^{\mathrm{phen}}(\vec{k}_M^\prime ; \vec{k}_M) =  Q_M \,
(k_M^\prime + k_M)_\nu\, f(Q) =
J_\nu^{\mathrm{point}}(\vec{k}_M^\prime ; \vec{k}_M)\, f(Q)
\end{equation}
multiplied with the covariant photon propagator $-g^{\mu\nu}/Q^2$
(and a kinematical factor),
\begin{eqnarray}\label{eq:1gexchamp}
\lefteqn{\langle v^\prime; \vec{k}_e^\prime, \mu_e^\prime;
\vec{k}_M^\prime \vert\, \hat{ V}_\mathrm{opt}(m)\, \vert v;
\vec{k}_e, \mu_e; \vec{k}_M \rangle_{\mathrm{on-shell}}}\nonumber\\
&=& v_0 \delta^3 (\vec{v}^{\, \prime} - \vec{v}\, )\, \frac{(2
\pi)^3 }{\sqrt{(\omega_{k_e^{\prime}}+\omega_{k_M^{\prime}})^3}
\sqrt{(\omega_{k_e^{\phantom{\prime}}}+\omega_{k_M^{\phantom{\prime}}})^3}}
\,j_\mu(\vec{k}_e^\prime, \mu_e^\prime;\vec{k}_e, \mu_e) \,
\frac{(-g^{\mu\nu})}{Q^2}\; J_\nu^{\mathrm{phen}}(\vec{k}_M^\prime ;
\vec{k_M})\, .\nonumber\\
\end{eqnarray}
Apart from the kinematical factor in front, the right-hand side of
Eq.~(\ref{eq:1gexchamp}) corresponds to the familiar one-photon
exchange amplitude for elastic electron-meson scattering
(calculated in the electron-meson center-of-mass system). It is
therefore justified to interpret $f(\Delta m)$ in
Eq.~(\ref{eq:mcurrphen}) as the electromagnetic meson form factor
which parameterizes the composite nature of the meson in a
phenomenological way.

At this point a warning is in order. At first sight one may get
the impression that the currents defined in Eqs.~(\ref{eq:elcurr})
and (\ref{eq:mcurrphen}) transform like 4-vectors. This is not
the case, since the momenta which appear in the currents are
 particle momenta in the center of mass of the
electron-meson system. The effect of a Lorentz transformation on
such momenta is just a Wigner rotation, as can be seen from
Eq.~(\ref{eq:vstateboost}). Currents with the correct covariance
properties, however, can be obtained by going back to the physical
particle momenta by means of a canonical boost with velocity
$\vec{v}$, i.e. $p_i^{(\prime)}=B_c(v)k_i^{(\prime)}$ with
$i=e,M$. If we define the electron and meson currents for the
physical particle momenta by
\begin{equation}
j_\mu (\vec{p}_e^{\,\prime}, \sigma_e^\prime;\vec{p}_e,
\sigma_e):= \left(B_c(v)\right)_\mu^{\phantom{\mu}\rho} j_\rho
(\vec{k}_e^{\,\prime}, \mu_e^\prime;\vec{k}_e, \mu_e) \,
D_{\mu_e^\prime \sigma_e^\prime}^{1/2\, \ast}
(R^{-1}_{\mathrm{W}}(\frac{k_e^\prime}{m_e},B_c(v))) \, D_{\mu_e
\sigma_e}^{1/2} (R^{-1}_{\mathrm{W}}(\frac{k_e}{m_e},B_c(v))) \,
\end{equation}
and
\begin{equation}
J_\nu^{\mathrm{phen}}(\vec{p}_M^{\,\prime} ; \vec{p}_M):=
\left(B_c(v)\right)_\nu^{\phantom{\nu}\rho}
J_\rho^{\mathrm{phen}}(\vec{k}_M^{\,\prime} ; \vec{k}_M)
\end{equation}
it can be checked that these currents exhibit the correct
covariance properties under Lorentz transformations and are
obviously conserved.

\subsection{\label{subsec:clustermass}Constituent level}
Our next objective will be to determine $f(\Delta m)$ starting from
a microscopic constituent-quark model for the meson. We will proceed
in a similar way as before with the only difference, that the mass
operator of our coupled channel system is now defined on a Hilbert
space which is the direct sum of channel Hilbert spaces consisting
of electron-quark-antiquark $\vert \psi_{e q \bar{q}} \rangle$ and
electron-quark-antiquark-photon states $\vert \psi_{e q \bar{q}
\gamma} \rangle$. Furthermore, the quark-antiquark pair is assumed
to be confined, so that the channel mass operators $\hat{M}_{e M}$
and $\hat{M}_{e M \gamma}$ in Eqs.~(\ref{eq:coupchann}) and
(\ref{eq:DynamicalEquationM}) become
\begin{equation}\label{eq:MeC}
\hat{M}_{e M} \rightarrow  \hat{M}_{e C} = \hat{M}_{e q \bar{q}}+
\hat{V}_{\mathrm{conf}}\, , \qquad \hat{M}_{e M \gamma} \rightarrow
\hat{M}_{e C \gamma} = \hat{M}_{e q \bar{q} \gamma} +
\hat{V}_{\mathrm{conf}}\, ,
\end{equation}
where $\hat{M}_{e q \bar{q}}$  and $\hat{M}_{e q \bar{q} \gamma}$
are free mass operators for the electron-quark-antiquark and
electron-quark-antiquark-photon systems and
$\hat{V}_{\mathrm{conf}}$ is an instantaneous confinement potential
acting between quark and antiquark. The subscript \lq\lq$C$\rq\rq\
of $\hat{M}_{e C}$ and $\hat{M}_{e C \gamma}$ should indicate that
the channel mass operators provide already a clustering of the
quark-antiquark subsystem. On the constituent level the vertex
operators $\hat{K}$ and $\hat{K}^\dag$ should describe the emission
and absorption of a photon by a pointlike (anti)quark. To
distinguish them from the hadronic case we will denote them by
$\hat{K}_{q\gamma}$ and $\hat{K}_{q\gamma}^\dag$.

On the constituent level it is advantageous to introduce two sets of
basis states. One set consists of velocity states that are
eigenstates of the free channel mass operators,
\begin{eqnarray}
\hat{M}_{e q \bar{q}} && \!\!\!\!\! \vert\,  v; \vec{k}_e, \mu_e;
\vec{k}_q, \mu_q; \vec{k}_{\bar{q}}, \mu_{\bar{q}} \rangle =
(\omega_{k_e} + \omega_{k_q} + \omega_{k_{\bar{q}}} ) \vert\,  v;
\vec{k}_e, \mu_e; \vec{k}_q, \mu_q; \vec{k}_{\bar{q}}, \mu_{\bar{q}}
\rangle \, ,
\nonumber\\
\hat{M}_{e q \bar{q} \gamma} && \!\!\!\!\!\vert\,  v; \vec{k}_e,
\mu_e; \vec{k}_q, \mu_q; \vec{k}_{\bar{q}}, \mu_{\bar{q}};
\vec{k}_{\gamma}, \mu_{\gamma} \rangle \\ & & = (\omega_{k_e} +
\omega_{k_q} + \omega_{k_{\bar{q}}} + \omega_{k_{\gamma}} ) \vert\,
v; \vec{k}_e, \mu_e; \vec{k}_q, \mu_q; \vec{k}_{\bar{q}},
\mu_{\bar{q}}; \vec{k}_{\gamma}, \mu_{\gamma} \rangle \, , \nonumber
\end{eqnarray}
the other set consists of velocity states that are eigenstates of
the channel mass operators with confinement potential
\begin{eqnarray}
\hat{M}_{e C} && \!\!\!\!\! \vert\,  \underline{v};
\vec{\underline{k}}_e, \underline{\mu}_e; \vec{\underline{k}}_C, n,
j,\tilde{m}_j, [\tilde{l}, \tilde{s} ] \rangle =
(\omega_{\underline{k}_e} + \omega_{\underline{k}_C} ) \vert\,
\underline{v}; \vec{\underline{k}}_e, \underline{\mu}_e;
\vec{\underline{k}}_C, n, j,\tilde{m}_j, [\tilde{l}, \tilde{s} ]
\rangle \, ,
\nonumber\\
\hat{M}_{e C \gamma} && \!\!\!\!\! \vert\,  \underline{v};
\vec{\underline{k}}_e, \underline{\mu}_e; \vec{\underline{k}}_C, n,
j,\tilde{m}_j, [\tilde{l}, \tilde{s} ]; \vec{\underline{k}}_\gamma,
\underline{\mu}_\gamma \rangle\\ & & = (\omega_{\underline{k}_e} +
\omega_{\underline{k}_C} + \omega_{\underline{k}_\gamma}) \vert\,
\underline{v}; \vec{\underline{k}}_e, \underline{\mu}_e;
\vec{\underline{k}}_C, n, j,\tilde{m}_j, [\tilde{l}, \tilde{s} ];
\vec{\underline{k}}_\gamma, \underline{\mu}_\gamma \rangle\, .
\nonumber
\end{eqnarray}
To distinguish these two sets of basis states we have underlined
velocities, momenta and spin projections for the velocity states of
$q\bar{q}$ clusters. $(n, j,\tilde{m}_j, [\tilde{l}, \tilde{s}])$
are the discrete quantum numbers of the $q\bar{q}$ cluster labeling
orbital excitation, total angular momentum, its projection, orbital
angular momentum, and total spin, respectively. The tilde should
indicate that the corresponding quantum numbers are defined in the
rest frame of the $q\bar{q}$ subsystem. Due to the occurrence of the
discrete quantum numbers $(n, j,\tilde{m}_j, [\tilde{l},
\tilde{s}])$ in cluster velocity states the orthogonality and
completeness relations, Eqs.~(\ref{eq:vnorm}) and (\ref{eq:vcompl}),
have to be supplemented by corresponding sums and Kronecker
deltas~\cite{Krassnigg:2003gh}.

The quantities that can be directly compared with
Eq.~(\ref{eq:voptmatr2}) are on-shell matrix elements of the
(constituent-level) optical potential
$\hat{V}_{\mathrm{opt}}^{\mathrm{const}}$ between cluster velocity
states with the cluster possessing the quantum numbers of the meson.
For a pseudoscalar meson one thus has to calculate
\begin{eqnarray}
\lefteqn{\langle  \underline{v}^\prime;
\vec{\underline{k}}_e^\prime, \underline{\mu}_e^\prime;
\vec{\underline{k}}_C^\prime, n, 0,0 , [0, 0] \vert\,
\hat{V}_{\mathrm{opt}}^{\mathrm{const}}(m) \vert\, \underline{v};
\vec{\underline{k}}_e, \underline{\mu}_e; \vec{\underline{k}}_C, n,
0,0, [0, 0] \rangle_{\mathrm{on-shell}}} \nonumber \\
&=& \langle  \underline{v}^\prime; \vec{\underline{k}}_e^\prime,
\underline{\mu}_e^\prime; \vec{\underline{k}}_C^\prime, n, 0,0 , [0,
0] \vert\, \hat{K}_{q\gamma}^\dag \left(\hat{M}_{e C \gamma}
-m\right)^{-1} \hat{K}_{q\gamma} \vert\, \underline{v};
\vec{\underline{k}}_e, \underline{\mu}_e; \vec{\underline{k}}_C, n,
0,0, [0, 0] \rangle\, ,
\end{eqnarray}
with $m=\omega_{\underline{k}_e}+\omega_{\underline{k}_C} =
\omega_{\underline{k}_e^\prime}+\omega_{\underline{k}_C^\prime}$ and
$\omega_{\underline{k}_e}=\omega_{\underline{k}_e^\prime}$ and
$\omega_{\underline{k}_C}=\omega_{\underline{k}_C^\prime}$. This can
again be done by inserting completeness relations for free and
cluster velocity states at appropriate places
\begin{eqnarray}\label{eq:voptclust}
\lefteqn{\langle  \underline{v}^\prime;
\vec{\underline{k}}_e^\prime, \underline{\mu}_e^\prime;
\vec{\underline{k}}_C^\prime, n, 0,0 , [0, 0] \vert\,
\hat{V}_{\mathrm{opt}}^{\mathrm{const}}(m) \vert\, \underline{v};
\vec{\underline{k}}_e, \underline{\mu}_e; \vec{\underline{k}}_C, n,
0,0, [0, 0] \rangle_{\mathrm{on-shell}}} \nonumber \\
&=& \langle  \underline{v}^\prime; \vec{\underline{k}}_e^\prime,
\underline{\mu}_e^\prime; \vec{\underline{k}}_C^\prime, n, 0,0 , [0,
0] \vert\, \mathds{1}_{e q \bar{q}}^{\prime}\,\hat{K}_{q\gamma}^\dag
\mathds{1}_{e q \bar{q} \gamma}^{\prime\prime\prime}\,
\left(\hat{M}_{e C \gamma}
-m\right)^{-1} \mathds{1}_{e C \gamma}^{\prime \prime}\, \nonumber\\
& &\times \mathds{1}_{e q \bar{q} \gamma}^{\prime \prime}\,
\hat{K}_{q\gamma} \mathds{1}_{e q \bar{q}}\, \vert\, \underline{v};
\vec{\underline{k}}_e, \underline{\mu}_e; \vec{\underline{k}}_C, n,
0,0, [0, 0] \rangle\, .
\end{eqnarray}
The matrix elements one needs to know are therefore typically scalar
products between free and cluster velocity states as well as matrix
elements of the vertex operators between free velocity states. With
our normalization of velocity states, cf. Eq.~(\ref{eq:vnorm}), such
matrix elements are given by (see also~\cite{Fuchsberger:2007,
Krassnigg:2003gh})
\begin{eqnarray}\label{eq:wf3}
\lefteqn{\langle v; \vec{k}_e, \mu_e; \vec{k}_q, \mu_q;
\vec{k}_{\bar{q}}, \mu_{\bar{q}} \vert\,  \underline{v};
\vec{\underline{k}}_e, \underline{\mu}_e; \vec{\underline{k}}_C, n,
j,\tilde{m}_j, [\tilde{l}, \tilde{s} ] \rangle} \nonumber\\ &=& (2
\pi)^{15/2}\, \underline{v}_0 \,
\delta^3(\vec{v}-\vec{\underline{v}})\,
\delta^3(\vec{k}_e-\vec{\underline{k}}_e)\,  \delta_{\mu_e
\underline{\mu}_e}\,  \sqrt{\frac{2 \omega_{\underline{k}_e} 2
\omega_{\underline{k}_C}}{(\omega_{\underline{k}_e} +
\omega_{\underline{k}_C})^3}} \sqrt{\frac{2 \omega_{k_e} 2
\omega_{k_{q\bar{q}}}}{(\omega_{k_e} + \omega_{k_{q\bar{q}}})^3}}
\sqrt{\frac{2 \omega_{\tilde{k}_q} 2
\omega_{\tilde{k}_{\bar{q}}}}{(\omega_{\tilde{k}_q} +
\omega_{\tilde{k}_{\bar{q}}})^3}} \nonumber\\
& &\times \sum_{\tilde{m}_l=-\tilde{l}}^{\tilde{l}}
\sum_{\tilde{m}_s=-\tilde{s}}^{\tilde{s}} \sum_{\tilde{\mu}_q ,\,
\tilde{\mu}_{\bar{q}}=\pm 1/2}\!\!
C_{\tilde{l}\tilde{m}_l\tilde{s}\tilde{m}_s}^{j\tilde{m}_j}
C_{\frac{1}{2}\tilde{\mu}_q
\frac{1}{2}\tilde{\mu}_{\bar{q}}}^{\tilde{s}\tilde{m}_s}\,
D_{\mu_q
\tilde{\mu}_q}^{1/2}(R_{\mathrm{W}}(\frac{\tilde{k}_q}{m_q},B_c(v_{q\bar{q}})))
\, D_{\mu_{\bar{q}}
\tilde{\mu}_{\bar{q}}}^{1/2}(R_{\mathrm{W}}(\frac{\tilde{k}_{\bar{q}}}{m_{\bar{q}}},
B_c(v_{q\bar{q}})))\nonumber\\ & &\times u_{n\tilde{l}}\,(\vert
\vec{\tilde{k}}_q\vert)\, Y_{\tilde{l}
\tilde{m}_l}(\hat{\tilde{k}}_q)\, ,
\end{eqnarray}
\begin{eqnarray}\label{eq:wf4}
\lefteqn{\langle v; \vec{k}_e, \mu_e; \vec{k}_q, \mu_q;
\vec{k}_{\bar{q}}, \mu_{\bar{q}};\vec{k}_\gamma, \mu_\gamma \vert\,
\underline{v}; \vec{\underline{k}}_e, \underline{\mu}_e;
\vec{\underline{k}}_C, n, j,\tilde{m}_j, [\tilde{l}, \tilde{s} ]
;\vec{\underline{k}}_\gamma, \underline{\mu}_\gamma\rangle} \nonumber\\
&=& (2 \pi)^{21/2}\, \underline{v}_0 \,
\delta^3(\vec{v}-\vec{\underline{v}})\,
\delta^3(\vec{k}_e-\vec{\underline{k}}_e)\,  \delta_{\mu_e
\underline{\mu}_e}\,
\delta^3(\vec{k}_\gamma-\vec{\underline{k}}_\gamma)\,
(-g_{\mu_\gamma \underline{\mu}_\gamma})\nonumber\\& & \times
\sqrt{\frac{2 \omega_{\underline{k}_e} 2 \omega_{\underline{k}_C} 2
\omega_{\underline{k}_\gamma}}{(\omega_{\underline{k}_e} +
\omega_{\underline{k}_C}+ \omega_{\underline{k}_\gamma})^3}}
\sqrt{\frac{2 \omega_{k_e} 2 \omega_{k_{q\bar{q}}} 2
\omega_{k_\gamma}}{(\omega_{k_e} + \omega_{k_{q\bar{q}}}+
\omega_{k_\gamma})^3}} \sqrt{\frac{2 \omega_{\tilde{k}_q} 2
\omega_{\tilde{k}_{\bar{q}}}}{(\omega_{\tilde{k}_q} +
\omega_{\tilde{k}_{\bar{q}}})^3}} \nonumber\\
& &\times \sum_{\tilde{m}_l=-\tilde{l}}^{\tilde{l}}
\sum_{\tilde{m}_s=-\tilde{s}}^{\tilde{s}} \sum_{\tilde{\mu}_q ,\,
\tilde{\mu}_{\bar{q}}=\pm 1/2}\!\!
C_{\tilde{l}\tilde{m}_l\tilde{s}\tilde{m}_s}^{j\tilde{m}_j}
C_{\frac{1}{2}\tilde{\mu}_q
\frac{1}{2}\tilde{\mu}_{\bar{q}}}^{\tilde{s}\tilde{m}_s}\,
D_{\mu_q
\tilde{\mu}_q}^{1/2}(R_{\mathrm{W}}(\frac{\tilde{k}_q}{m_q},B(v_{q\bar{q}})))
\, D_{\mu_{\bar{q}}
\tilde{\mu}_{\bar{q}}}^{1/2}(R_{\mathrm{W}}(\frac{\tilde{k}_{\bar{q}}}{m_{\bar{q}}},
B(v_{q\bar{q}})))\nonumber\\ & &\times u_{n\tilde{l}}\,(\vert
\vec{\tilde{k}}_q\vert)\, Y_{\tilde{l}
\tilde{m}_l}(\hat{\tilde{k}}_q)\, ,
\end{eqnarray}\label{eq:qvertex}
\begin{eqnarray} \lefteqn{
\langle v^\prime; \vec{k}_e^\prime, \mu_e^\prime; \vec{k}_q^\prime,
\mu_q^\prime; \vec{k}_{\bar{q}}^\prime, \mu_{\bar{q}}^\prime;
\vec{k}_\gamma^\prime, \mu_\gamma^\prime \vert \,\hat{K}\, \vert v;
\vec{k}_e, \mu_e; \vec{k}_q, \mu_q; \vec{k}_{\bar{q}}, \mu_{\bar{q}}
\rangle}
\nonumber\\
&=& v_0 \, \delta^3(\vec{v}^\prime-\vec{v})\,
\frac{(2\pi)^3}{\sqrt{(\omega_{k_e^\prime}+\omega_{k_q^\prime}+
\omega_{k_{\bar{q}}^\prime}+ \omega_{k_{\gamma}^\prime})^3}
\sqrt{(\omega_{k_e}+\omega_{k_q^{\phantom{\prime}}}
+\omega_{k_{\bar{q}}})^3}} \nonumber
\\ & &\times  \langle v^\prime; \vec{k}_e^\prime, \mu_e^\prime; \vec{k}_q^\prime,
\mu_q^\prime; \vec{k}_{\bar{q}}^\prime, \mu_{\bar{q}}^\prime;
\vec{k}_\gamma^\prime, \mu_\gamma^\prime \vert \,\left(
\hat{\mathcal{L}}_{\mathrm{int}}^{e \gamma}(0) +
\hat{\mathcal{L}}_{\mathrm{int}}^{q \gamma}(0) \right) \vert v;
\vec{k}_e, \mu_e; \vec{k}_q, \mu_q; \vec{k}_{\bar{q}}, \mu_{\bar{q}}
\rangle \nonumber\\
&=& v_0 \, \delta^3(\vec{v}^\prime-\vec{v})\,
\frac{(2\pi)^3}{\sqrt{(\omega_{k_e^\prime}+\omega_{k_M^\prime}+
\omega_{k_{\gamma}^\prime})^3}
\sqrt{(\omega_{k_e}+\omega_{k_M^{\phantom{\prime}}})^3}}(-1)\nonumber\\
&&\times \left[ \, Q_e\,
\bar{u}_{\mu_e^\prime}(\vec{k}_e^\prime)\gamma_\nu
u_{\mu_e}(\vec{k}_e)\,
\epsilon^\nu(\vec{k}_{\gamma}^\prime,\mu_{\gamma}^\prime)\, (2
\pi)^3 2 \omega_{k_q} \delta^3(\vec{k}_q^\prime - \vec{k}_q) \, (2
\pi)^3 2 \omega_{k_{\bar{q}}} \delta^3(\vec{k}_{\bar{q}}^\prime -
\vec{k}_{\bar{q}})\right. \\
& & \left. \;\; + \, Q_q\,
\bar{u}_{\mu_q^\prime}(\vec{k}_q^\prime)\gamma_\nu
u_{\mu_q}(\vec{k}_q)\,
\epsilon^\nu(\vec{k}_{\gamma}^\prime,\mu_{\gamma}^\prime)\, (2
\pi)^3 2 \omega_{k_e} \delta^3(\vec{k}_e^\prime - \vec{k}_e) \, (2
\pi)^3 2 \omega_{k_{\bar{q}}} \delta^3(\vec{k}_{\bar{q}}^\prime -
\vec{k}_{\bar{q}}) \right. \nonumber \\
& & \left. \;\; + \, Q_{\bar{q}}\,
\bar{v}_{\mu_{\bar{q}}}(\vec{k}_{\bar{q}})\gamma_\nu
v_{\mu_{\bar{q}}^\prime}(\vec{k}_{\bar{q}})\,
\epsilon^\nu(\vec{k}_{\gamma}^\prime,\mu_{\gamma}^\prime)\, (2
\pi)^3 2 \omega_{k_e} \delta^3(\vec{k}_e^\prime - \vec{k}_e) \, (2
\pi)^3 2 \omega_{k_q} \delta^3(\vec{k}_q^\prime - \vec{k}_q)
\right]\, , \nonumber
\end{eqnarray}
and the hermitian conjugate expressions, respectively.
$\hat{\mathcal{L}}_{\mathrm{int}}^{q \gamma}(x)$ is again the
usual interaction density for spinor quantum electrodynamics
describing the coupling of the photon to a quark or antiquark.
Spins and angular momenta are coupled in the rest frame of the
$q\bar{q}$ subsystem. This is the reason that the Wigner $D$
functions $D_{\mu \tilde{\mu}}^{1/2}(R_W)$ relating the spin
projection of the (anti)quark in the $q\bar{q}$ rest frame to the
spin projection in the overall $eq\bar{q}$($\gamma$)
center-of-mass frame appear in Eqs.~(\ref{eq:wf3}) and
(\ref{eq:wf4}). The pertinent Wigner rotation is given by
Eq.~(\ref{eq:wignerrot}) with
\begin{equation}
v_{q\bar{q}}=\frac{k_{q\bar{q}}}{m_{q\bar{q}}}=\frac{k_q +
k_{\bar{q}}}{m_{q\bar{q}}}\quad \mathrm{and}\quad
m_{q\bar{q}}=\tilde{k}_q^0+\tilde{k}_{\bar{q}}^0 =
\sqrt{(k_q^0+k_{\bar{q}}^0)^2-(\vec{k}_q+\vec{k}_{\bar{q}})^2}\, .
\end{equation}
For later purposes we note that
$\vec{k}_e+\vec{k}_q+\vec{k}_{\bar{q}}=\vec{k}_e+\vec{\underline{k}}_C=0$
and hence $\vec{k}_q+\vec{k}_{\bar{q}}=\vec{\underline{k}}_C$ so
that $\vec{v}_{q\bar{q}}=\vec{\underline{k}}_C/m_{q\bar{q}}$. The
$u_{n\tilde{l}}\,(\vert \vec{\tilde{k}}_q\vert)$ form a complete
set of radial wave functions for the confined quark-antiquark pair
orthonormalized according to
\begin{equation}
\int_0^\infty d\tilde{k}\,\tilde{k}^2\, u_{n^\prime
\tilde{l}^\prime}(\tilde{k}) u_{n\tilde{l}}(\tilde{k}) =
\delta_{n^\prime n}\, \delta_{\tilde{l}^\prime \tilde{l}}\, .
\end{equation}
The $Y_{\tilde{l} \tilde{m}_l}(\hat{\tilde{k}})$ are usual spherical
harmonics depending on $\hat{\tilde{k}}=\vec{\tilde{k}}/\vert
\vec{\tilde{k}} \vert$. In Eqs.~(\ref{eq:wf3}) and (\ref{eq:wf4}) it
has been assumed for simplicity that the confining potential is spin
independent.

\begin{figure}
\includegraphics{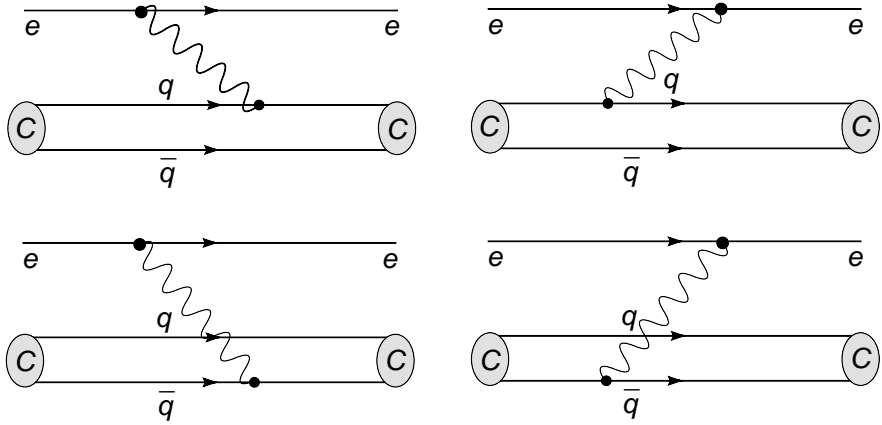}% Here is how to import EPS art
\caption{\label{fig:feyn} Graphical representation of the four
contributions to the one-photon-exchange optical potential for the
scattering of an electron by a quark-antiquark cluster.}
\end{figure}
With these results for the matrix elements most of the sums and
integrals on the right-hand side of Eq.~(\ref{eq:voptclust}) (coming
from the inserted completeness relations) can be carried out analytically
by means of Dirac and Kronecker deltas. If self-energy contributions
are again neglected the optical potential consists of four terms
which are depicted in Fig.~\ref{fig:feyn}. Here the blobs which
connect the quark and antiquark lines symbolize integrals over wave
functions of the incoming and outgoing quark-antiquark cluster.
These integrals are the same for both time orderings, i.e. for the
absorption and emission of the photon by the (anti)quark. Thus both
time orderings can be combined, as on the hadronic level, and one is
left with a photon-exchange contribution for the quark and another
for the antiquark. If, in addition, quark and antiquark have the
same mass $m_q = m_{\bar{q}}$, even the integrals for the quark and
antiquark contributions become the same so that the final expression
for the one-photon-exchange optical potential takes on a simple
form on the constituent level comparable to that on the hadronic
level (cf. Eq.~(\ref{eq:1gexchamp})):
\begin{eqnarray}\label{eq:voptcluster2}
\lefteqn{\langle  \underline{v}^\prime;
\vec{\underline{k}}_e^\prime, \underline{\mu}_e^\prime;
\vec{\underline{k}}_C^\prime, n, 0,0 , [0, 0] \vert\,
\hat{V}_{\mathrm{opt}}^{\mathrm{const}}(m) \vert\, \underline{v};
\vec{\underline{k}}_e, \underline{\mu}_e; \vec{\underline{k}}_C, n,
0,0, [0, 0] \rangle_{\mathrm{on-shell}}}\nonumber\\
&=& \underline{v}_0 \delta^3 (\vec{\underline{v}}^{\, \prime} -
\vec{\underline{v}}\, )\, \frac{(2 \pi)^3
}{\sqrt{(\omega_{\underline{k}_e^{\prime}}+
\omega_{\underline{k}_C^{\prime}})^3}
\sqrt{(\omega_{\underline{k}_e^{\phantom{\prime}}}+
\omega_{\underline{k}_C^{\phantom{\prime}}})^3}}
\,j_\mu(\vec{\underline{k}}_e^\prime,
\underline{\mu}_e^\prime;\vec{\underline{k}}_e, \underline{\mu}_e)
\, \frac{(-g^{\mu\nu})}{Q^2}\;
J_\nu^{\mathrm{micro}}(\vec{\underline{k}}_C^\prime ;
\vec{\underline{k}}_C)\, ,\nonumber\\
\end{eqnarray}
with the (preliminary) microscopic meson current being given by
\begin{eqnarray}\label{eq:mcurrmicro}
\lefteqn{J_\nu^{\mathrm{micro}}(\vec{\underline{k}}_C^\prime ;
\vec{\underline{k}}_C) = (Q_q+Q_{\bar{q}})
\sqrt{\omega_{\underline{k}_C} \omega_{\underline{k}_C^{\prime}}}
\int\, \frac{d^3\tilde{k}_q^\prime}{\omega_{k_q}}\,
\sqrt{\frac{\omega_{\tilde{k}_q}}{\omega_{\tilde{k}_q^\prime}}} \,
\sqrt{\frac{\omega_{k_{q\bar{q}}}}{\omega_{k_{q\bar{q}}^\prime}}}}\nonumber\\
&& \times  u_{n 0}^\ast\,(\vert \vec{\tilde{k}}_q^\prime\vert)\,
Y_{0 0}^\ast(\hat{\tilde{k}}_q^\prime)\, u_{n 0}\,(\vert
\vec{\tilde{k}}_q\vert)\, Y_{0 0}(\hat{\tilde{k}}_q)
\bigg[\sum_{\mu_q,\mu_q^\prime =\pm \frac{1}{2}}
\bar{u}_{\mu_q^\prime}(\vec{k}_q^\prime)\gamma_\nu
u_{\mu_q}(\vec{k}_q) \\ &&\times D^{1/2}_{\mu_q\mu_q^\prime}
\left( R_{\mathrm{W}}(\frac{\tilde{k}_q}{m_q},
B_c(v_{q\bar{q}}))\,
R^{-1}_{\mathrm{W}}(\frac{\tilde{k}_{\bar{q}}}{m_{\bar{q}}},
B_c(v_{q\bar{q}}))\,
R_{\mathrm{W}}(\frac{\tilde{k}_q^\prime}{m_q},
B_c(v_{q\bar{q}}^\prime))\,
R^{-1}_{\mathrm{W}}(\frac{\tilde{k}_{\bar{q}}^\prime}{m_{\bar{q}}},
B_c(v_{q\bar{q}}^\prime))\right) \bigg]\, . \nonumber
\end{eqnarray}
According to our notation introduced previously, momenta without
and with tilde satisfy $\vec{k}_e+\vec{k}_q+\vec{k}_{\bar{q}}=0$
and $\vec{\tilde{k}}_q+\vec{\tilde{k}}_{\bar{q}}=0$, respectively.
These two sets of momenta are connected via the canonical spin
boost $B_c(v_{q\bar{q}})$, e.g. $B_c(v_{q\bar{q}}) \tilde{k}_q =
k_q$. An analogous property holds for the primed momenta. Momenta
with and without prime are are related by $\vec{k}_q^{\, \prime} =
\vec{k}_q + (\vec{\underline{k}}_C^\prime - \vec{\underline{k}}_C)
= \vec{k}_q + \vec{\underline{k}}_\gamma = \vec{k}_q +
\vec{k}_\gamma$. This means that we have 3-momentum conservation
at the photon-quark vertex, a property which one would not expect
in point-form quantum mechanics. Here it results from the
velocity-state representation, in particular the associated
center-of-mass kinematics, and the spectator conditions for
electron and antiquark. For the physical momenta, i.e. the
center-of-mass momenta boosted by $B_c(v)$, none of the 4-momentum
components is conserved at the electromagnetic vertices, in
general (if $v \neq 0$). As we have seen already on the hadronic
level, the denominator of the (covariant) photon propagator (which
includes both time orderings) is given by $Q^2=-\underline{q}_\mu
\underline{q}^\mu$ with
$\underline{q}^\mu=(\underline{k}_C^\prime-\underline{k}_C)^\mu$
being the 4-momentum transfer between incoming and outgoing
cluster \footnote{Note that $\vec{q}=\vec{k}_\gamma$, but $q^0
\neq k_\gamma^0 = \omega_{k_\gamma}=\vert \vec{k}_\gamma \vert$.}.
The Wigner $D$ function occurring in Eq.~(\ref{eq:mcurrmicro})
results from combining individual Wigner $D$ functions for the
quark and antiquark. Thereby the sums over the spin projections
can be carried out with the help of the spectator conditions and
the Clebsch-Gordan coefficients $C^{00}_{\frac12\tilde{\mu}_q
\frac12\tilde{\mu}_{\bar{q}}}$ which couple the quark and
antiquark spins to zero meson spin.

The warning that we have sounded at the end of the previous
subsection has to be repeated here. The microscopic meson
current, as defined in Eq.~(\ref{eq:mcurrmicro}) does not behave
like a 4-vector under Lorentz transformations $\Lambda$. Rather it
transforms by the Wigner rotation $R_{\mathrm
W}(v,\Lambda)$. The current with the correct transformation
properties is again the one involving the physical meson momenta,
i.e.
\begin{equation}\label{eq:jcov}
J_\nu^{\mathrm{micro}}(\vec{p}_C^{\,\prime} ; \vec{p}_C):=
\left(B_c(v)\right)_\nu^{\phantom{\nu}\rho}
J_\rho^{\mathrm{micro}}(\vec{\underline{k}}_C^{\,\prime} ;
\vec{\underline{k}}_C)\, .
\end{equation}
In general, the 4-momentum transfer between incoming and outgoing
(active) quark $q^\mu=(k_q^\prime-k_q)^\mu$ deviates from the
4-momentum transfer between incoming and outgoing cluster
$\underline{q}^\mu=(\underline{k}_C^\prime-\underline{k}_C)^\mu$.
Whereas the 3-momentum transfers are the same on the hadronic and
on the constituent level, i.e.
$\vec{\underline{q}}=\vec{\underline{k}}_C^\prime-
\vec{\underline{k}}_C= \vec{k}_q^\prime-\vec{k}_q=\vec{q}$, the
zero components differ, $\underline{q}^0\neq q^0$. Due to the
center-of-mass kinematics associated with the velocity states
$\omega_{\underline{k}_C}=\omega_{\underline{k}_C^\prime}$ and
hence $\underline{q}^0=0$, but on the other hand
$\omega_{k_q^\prime}^{2}=\omega_{k_q}^2-2 \vec{q}\cdot
\vec{k}_{\bar{q}}^{\, \prime}\,$ so that $q^0\neq 0$. This means
that not all of the 4-momentum transferred to the cluster is also
transferred to the active constituent. Nevertheless, the
microscopic current
$J_\nu^{\mathrm{micro}}(\vec{\underline{k}}_C^{\,\prime} ;
\vec{\underline{k}}_C)$ and hence also
$J_\nu^{\mathrm{micro}}(\vec{p}_C^{\,\prime} ; \vec{p}_C)$ is
 conserved, i.e. $(p_C^\prime - p_C)^\nu
J_\nu^{\mathrm{micro}}(\vec{p}_C^{\,\prime} ; \vec{p}_C)=0$. The
analytical proof of current conservation amounts to showing that
the integral in Eq.~(\ref{eq:mcurrmicro}) vanishes if the spinor
product $\bar{u}\gamma_\nu u$ is replaced by $\bar{u}\gamma_0
(\omega^\prime_q-\omega_q) u$. By a change of the integration
variables $d^3\tilde{k}_q^\prime\rightarrow d^3\tilde{k}_q$ it can
be shown that the integral over $\bar{u}\gamma_0 \omega^\prime_q
u$ goes over into the integral over $\bar{u}\gamma_0 \omega_q u$
and vice versa so that the difference vanishes.
%
%, as we have checked numerically (but have not been able to show
%analytically).

\section{\label{sec:formfac}Identifying the meson form factor}
By comparing the optical one-photon-exchange potentials for
electron-meson scattering on the hadronic and the constituent
levels we are now in the position to extract the vertex form
factor in a unique way. If the cluster has the same (external)
quantum numbers as the meson, we can equate the right-hand-sides
of Eqs.~(\ref{eq:1gexchamp}) and (\ref{eq:voptcluster2}) to find
\begin{equation}\label{eq:ffprelim}
f(\Delta m, \vert \vec{k}_M \vert) = \frac{j^\mu(\vec{k}_e^\prime,
{\mu}_e^\prime;\vec{k}_e, {\mu}_e)
J_\mu^{\mathrm{micro}}(\vec{k}_M^\prime ;
\vec{k}_M)}{j^\mu(\vec{k}_e^\prime, {\mu}_e^\prime;\vec{k}_e,
{\mu}_e) J_\mu^{\mathrm{point}}(\vec{k}_M^\prime ; \vec{k}_M)}\, ,
\end{equation}
with $J_\mu^{\mathrm{point}}$ being defined in
Eq.~(\ref{eq:mcurrphen}). Here we have introduced $\vert \vec{k}_M
\vert$ as a further argument of the vertex form factor since it is
a priori not clear from its microscopic expression,
Eq.~(\ref{eq:ffprelim}), that it will only depend on $\Delta m$.
Poincar\'e invariance of our Bakamjian-Thomas type approach would
not be spoiled by vertex form factors that also depend on the whole
set of independent Lorentz invariants involved in the process. For
elastic electron-meson scattering these are, e.g., the Mandelstam
variables $s=\left(\sqrt{m_e+\vert \vec{k}_M
\vert^2}+\sqrt{m_M+\vert \vec{k}_M \vert^2}\right)^2$ and $t=-\Delta
m^2=-Q^2$.

\begin{figure}
\begin{center}
\includegraphics[clip=7cm,width=10cm]{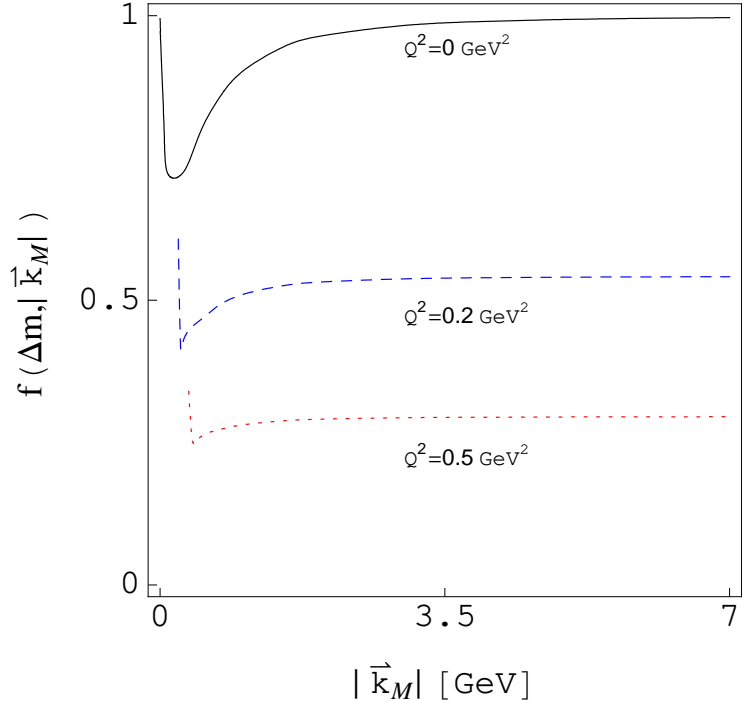}
 \caption{\label{fig:kdep}$\vert \vec{k}_M\vert$-dependence of the
pion form factor for different values of $Q^2=\Delta m^2$
($Q^2=0$~GeV$^2$ solid, $Q^2=0.2$~GeV$^2$ dashed, $Q^2=0.5$~GeV$^2$
dotted) as evaluated by means of Eq.~(\ref{eq:ffprelim}).}
 \end{center}
 \end{figure}
A reasonable microscopic model for an electromagnetic hadron form
factor should, of course, only depend on the momentum transfer
$\Delta m = \sqrt{-t}$ and not on $\sqrt{s}$, the invariant mass of
the electron-meson system. To check, whether this is the case for
our form factor expression, Eq.~(\ref{eq:ffprelim}), we make a
numerical study with a simple harmonic oscillator wave function for
the quark-antiquark bound state:
\begin{equation}\label{eq:u00}
u_{00}(\vert \vec{\tilde{k}}_q\vert) = \frac{2}{\pi^{1/4}
a^{3/2}}\mathrm{e}^{-\frac{\vert \vec{\tilde{k}}_q\vert^2}{2 a^2}}\,
.
\end{equation}
There are just two free parameters in this simple model, the
oscillator parameter $a$ and the constituent-quark mass $m_q$. For
later comparison we take these parameters from a paper of Chung,
Coester, and Polyzou~\cite{Chung:1988mu} in which a reasonable fit
of the pion form-factor data has been achieved using front-form
quantum mechanics by choosing $a=0.35$~GeV$^2$ and $m_q=0.21$~GeV.
We want to emphasize that our goal is not to end up with an
optimal fit of the pion form factor, but we rather want to exhibit
the virtues of our point-form approach and compare it to other
attempts to develop microscopic models for hadron form factors.
The dependence of the pion form factor on $\vert \vec{k}_M\vert$
(or Mandelstam $s$) for different values of the momentum transfer
$Q$, as evaluated by means of Eq.~(\ref{eq:ffprelim}), is shown in
Fig.~\ref{fig:kdep}. What one observes is a moderate dependence on
$\vert \vec{k}_M\vert$ at low energies which vanishes nearly
completely for $\vert \vec{k}_M\vert
> 2$~GeV. In addition, this $\vert \vec{k}_M\vert$-dependence
becomes weaker with increasing momentum transfer $Q^2$.

These observations mean that the electromagnetic meson current
$J_\nu^{\mathrm{micro}}(\vec{{p}}_M^{\, \prime} ; \vec{{p}}_M)$
(cf. Eq.~(\ref{eq:jcov})), which we have derived from the
one-photon exchange optical potential, does not have all the
properties it should have. It satisfies current conservation and
transforms like a 4-vector, but it cannot be written as a sum of
covariants times Lorentz invariants ($Q^2$ and $m_M^2$) which are
solely built from the incoming and outgoing meson 4-momenta
($p_M^\mu$ and $p_M^{\mu\,\prime}$). Actually, this result is not
too surprising. Recall that the electromagnetic vertex has been
approximated such that electron-meson scattering could be treated
within the Bakamjian-Thomas framework. Associated with the
Bakamjian-Thomas framework is the problem of cluster
separabilty~\cite{Keister:1991sb}. In our approach this problem is
inherent in the definition of the cluster wave functions via
velocity states, Eqs.~(\ref{eq:wf3}) and (\ref{eq:wf4}). The wave
function of the cluster changes with the presence of additional
particles, even if these particles don't interact with the
cluster. This change is essentially proportional to the fraction
of the cluster binding energy over the invariant mass of the whole
electron-meson system.

Therefore the $\vert \vec{k}_M\vert$-dependence vanishes rather
quickly with increasing invariant mass and it is suggestive to take
the limit $\vert \vec{k}_M\vert \rightarrow \infty$. In this way
one ends up with a vertex form factor that depends only on the
momentum transfer $Q^2$. It is even more interesting to see
explicitly that in this limit the microscopic meson current,
Eq.~(\ref{eq:mcurrmicro}), can be cast into the same form as the
phenomenological meson current, Eq.~(\ref{eq:mcurrphen}), i.e.
\begin{equation}\label{eq:jmmicro}
J_\nu^{\mathrm{micro}}(\vec{{k}}_M^\prime ; \vec{{k}}_M)
\stackrel{\vert \vec{k}_M\vert \rightarrow \infty}{\longrightarrow}
(Q_q+Q_{\bar{q}}) \left( k_M^\prime + k_M\right)_\nu F(Q^2)\, .
\end{equation}
This allows us to extract the electromagnetic form factor directly
from the limiting expression of
$J_\nu^{\mathrm{micro}}(\vec{{k}}_M^\prime ; \vec{{k}}_M)$ without
making use of Eq.~(\ref{eq:ffprelim}). The final result is
\begin{equation}\label{eq:formfactor}
F(Q^2) = \lim_{\vert \vec{ k}_M\vert\rightarrow \infty} f(\Delta m =
Q, \vert \vec{ k}_M\vert ) =\int\mathrm{d}^3\tilde{k}^\prime_q
\sqrt{\frac{m_{q\bar{q}}}{m'_{q\bar{q}}}}\, \mathcal{S}\, u_{n
0}^\ast\,(\vert \vec{\tilde{k}}_q^\prime\vert)\, Y_{0
0}^\ast(\hat{\tilde{k}}_q^\prime)\, u_{n 0}\,(\vert
\vec{\tilde{k}}_q\vert)\, Y_{0 0}(\hat{\tilde{k}}_q)\, .
\end{equation}
The spin-rotation factor $\mathcal S$ is the $\vert \vec{k}_M\vert
\rightarrow \infty$ limit of the trace of the Wigner $D$ function
occurring in Eq.~(\ref{eq:mcurrmicro}). In order to find explicit
expressions for $\vec{\tilde{k}}_q$, $m_{q\bar{q}}$ and $\mathcal
S$ in terms of the integration variable $\vec{\tilde{k}}_q^\prime$
and the momentum transfer $Q$ we have to specify our kinematics.
For convenience we choose the (1,3)-plane as the scattering plane
and the 3-momenta such that
\begin{equation}\label{eq:kM}
\vec{k}_M=\left(%
\begin{array}{c} -\frac{Q}{2}\\0\\
\sqrt{\vec{k}_M^2-\frac{Q^2}{4}}
\end{array}
\right)\, , \qquad
\vec{q}=\left(%
\begin{array}{c} Q\\0\\
0
\end{array}
\right)\, , \qquad \vec{k}_M^\prime = \vec{k}_M + \vec{q}\, .
\end{equation}
It should be noted that $Q$ is restricted by $Q<2\vert
\vec{k}_M\vert$. With this choice of kinematics we
find~\cite{Fuchsberger:2007}
\begin{equation}\label{eq:kinf}
\vec{\tilde{k}}_q\stackrel{\vert \vec{k}_M\vert \rightarrow
\infty}{\longrightarrow} \left(%
\begin{array}{c} \tilde{k}_q^{1\prime}+\left(
\frac{\tilde{k}_q^{3\prime}}{m_{q\bar{q}}^{\prime }}-\frac12 \right)Q \\
\tilde{k}_q^{2\prime}\\ \tilde{k}_q^{3\prime}\frac{m_{q\bar{q}}}{m_{q\bar{q}}^{\prime }}
\end{array}
\right)\, ,
 \end{equation}
\medskip
\begin{equation}
m_{q\bar{q}}\stackrel{\vert \vec{k}_M\vert \rightarrow
\infty}{\longrightarrow} \sqrt{m_{q\bar{q}}^{\prime
2}-\frac{4\tilde{k}_q^{1\prime}
m_{q\bar{q}}^{\prime}}{2\tilde{k}_q^{3\prime}+m_{q\bar{q}}^{\prime}}Q+\frac{\left(
m_{q\bar{q}}^{\prime}-2\tilde{k}_q^{3\prime}\right)
}{2\tilde{k}_q^{3\prime}+m_{q\bar{q}}^{\prime}}Q^2}\, ,
\end{equation}
and
\begin{equation}\label{eq:spinrot}
\mathcal{S} = \frac{m_{q\bar{q}}^\prime}{m_{q\bar{q}}}-\frac{2
\tilde{k}_q^{\prime 1} \, Q}{m_{q\bar{q}} (m_{q\bar{q}}^\prime + 2
\tilde{k}_q^{\prime 3})}\, .
\end{equation}
$F(Q^2)$ is thus independent of the reference frame. The integrand
on the right-hand side of Eq.~(\ref{eq:formfactor}) depends only
on the momentum transfer $Q$ and the internal (anti)quark momentum
$\vec{\tilde k}^\prime$, which is integrated over. Details of the
confinement dynamics enter solely via the form of the bound-state
wave function $u_{n 0}\,(\vert \vec{\tilde{k}}_q\vert)\, Y_{0
0}(\hat{\tilde{k}}_q)$ and not via the mass of the bound
$q\bar{q}$ cluster. What we have achieved with
Eq.~(\ref{eq:formfactor}) is an impulse approximation to the
electromagnetic meson form factor. In the limit $\vert
\vec{k}_M\vert \rightarrow \infty$ the whole photon momentum is
transferred to one of the constituents (since $q^\mu
\stackrel{\vert \vec{k}_M\vert \rightarrow
\infty}{\longrightarrow} \underline{q}^\mu$), whereas the other
one acts as a spectator. Furthermore it is important to note that
the electromagnetic meson current (expressed in terms of physical
particle momenta) $J_\nu^{\mathrm{micro}}(\vec{{p}}_M^{\, \prime}
; \vec{{p}}_M)$ acquires the correct continuity, covariance and
cluster-separability properties in the limit $\vert \vec{k}_M\vert
\rightarrow \infty$, as can directly be seen from boosting
Eq.~(\ref{eq:jmmicro}) by $B_c(v)$. We can also turn this around
and say that we have found a reference frame for the $\gamma^\ast
M \rightarrow M$ subprocess in which a one-body constituent
current
 already provides the correct continuity, covariance and
cluster-separability properties for the electromagnetic meson
current.

\begin{figure}
\begin{center}
\includegraphics[clip=7cm,width=10cm]{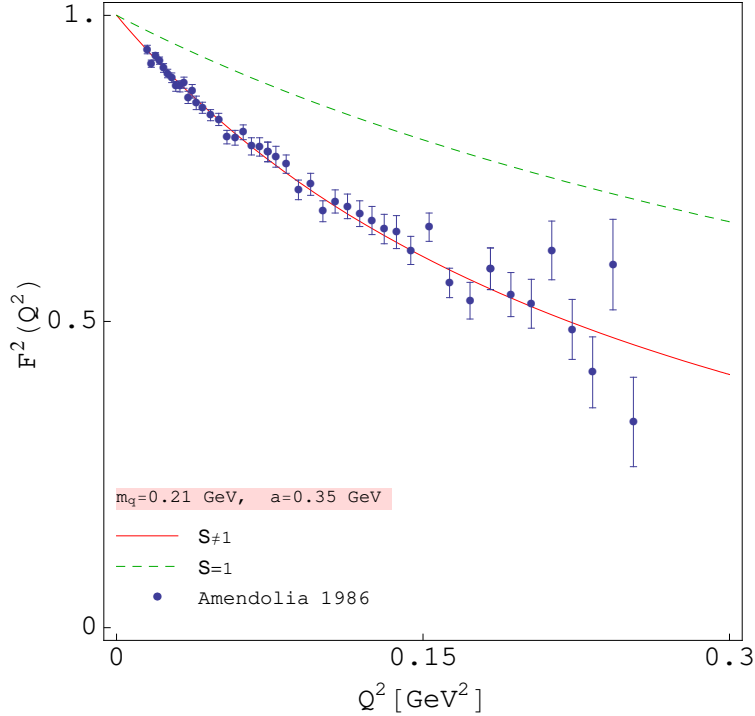}
 \caption{\label{fig:fflowQ2} $Q^2$ dependence of the pion
 form factor in the low-$Q^2$ region, as evaluated by means of
 Eq.~(\ref{eq:formfactor}) with (solid) and without (dashed)
 spin-rotation factor $\mathcal{S}$. Data are taken from
 Ref.~\cite{Amendolia:1986wj}}
 \end{center}
\end{figure}
\begin{figure}
\begin{center}
\includegraphics[clip=7cm,width=10cm]{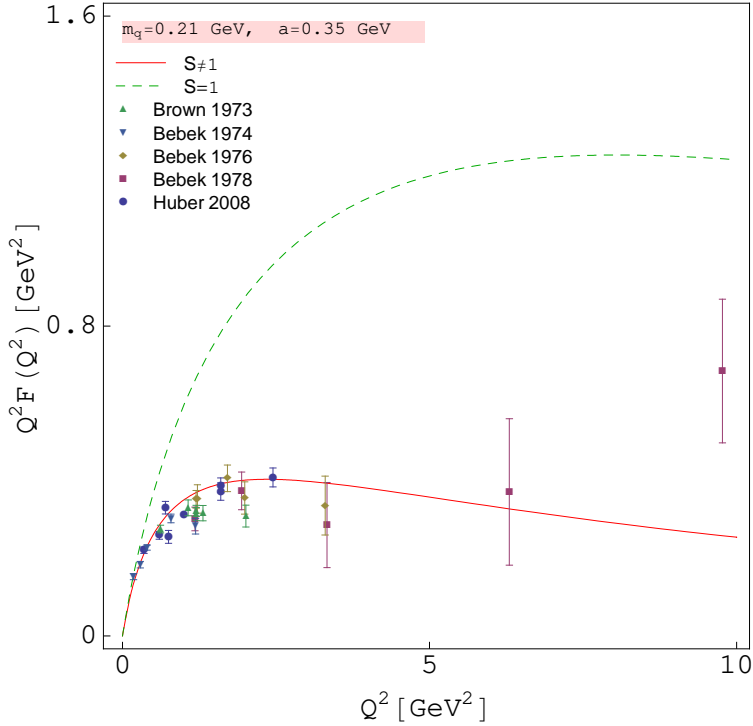}
 \caption{\label{fig:ffhighQ2} $Q^2$ dependence of the
 pion form factor scaled by $Q^2$, as evaluated by means of
 Eq.~(\ref{eq:formfactor}) with (solid) and without (dashed)
 spin-rotation factor $\mathcal{S}$. Data are taken from
 Refs.~\cite{Brown:1973wr,Bebek:1974a,Bebek:1974b,Bebek:1977pe,Huber:2008id}}
 \end{center}
 \end{figure}
Numerical results obtained with our final expression for the
electromagnetic pion form factor, Eq.~(\ref{eq:formfactor}), and the
simple harmonic-oscillator wave function  introduced above
are depicted in Figs.~\ref{fig:fflowQ2} and \ref{fig:ffhighQ2} along
with experimental data. The one-body constituent current, which we
end up with, together with a simple two-parameter
harmonic-oscillator wave function is seen to provide  a
reasonable fit to the pion electromagnetic form factor.
Also shown is  the role of the spin-rotation factor $\mathcal{S}$.
The quark spin obviously has a substantial effect on the
electromagnetic form factor over nearly the whole momentum-transfer
range and thus  cannot be neglected. As will become clear in the
following all these findings are not too surprising. They coincide
with corresponding statements made in Ref.~\cite{Chung:1988mu} (from
where we have taken the wave function parametrization).

\section{\label{sec:results}Comparison with other approaches}

\subsection{\label{subsec:frontform}Comparison with front-form results}
The fact that we extract the electromagnetic meson form factor in
the limit $\vert \vec{k}_M\vert \rightarrow \infty$ means that the
$\gamma^\ast M \rightarrow M$ subprocess is considered in the
infinite momentum frame of the meson. This offers  the
possibility of making a direct comparison with form factor analyses
using front form of relativistic dynamics. The $\vert
\vec{k}_M\vert \rightarrow \infty$ limit of the kinematics chosen
in Eq.~(\ref{eq:kM}) implies in particular that we work in a
reference frame in which  the plus component of the 4-momentum
transfer $q^\mu=(k_M^\prime-k_M)^\mu$ vanishes. $q^{+}=0$ frames
are also popular for form-factor studies in front
form~\cite{Keister:1991sb, Chung:1988mu, Coester:2005cv,
Simula:2002vm}. One reason is that the impulse approximation can
be formulated consistently in any $q^+=0$ frame (for the plus
component of the current operator)~\cite{Keister:1991sb}. The
other reason is that so called \lq\lq Z-graphs\rq\rq\ are
suppressed in such frames~\cite{Simula:2002vm}.

Our point form calculation can be related to front form results by
an appropriate change of variables. To show this we define the
longitudinal momentum fractions $\xi$ and $\xi^\prime$ as
\begin{equation}
\xi^{(\prime)}-\frac{1}{2}=\frac{\tilde{k}_q^{3(\prime)}}{m_{q\bar{q}}^{(\prime)
}}
\end{equation}
and introduce the short-hand notation
\begin{equation}
\kappa_\perp^{(\prime)}=\left(
\begin{array}{c}\tilde{k}_q^{1(\prime)}\\\tilde{k}_q^{2(\prime)}
\end{array}\right)
\end{equation}
for the intrinsic transverse momentum of the active incoming
(outgoing) quark. From Eq.~(\ref{eq:kinf}) we infer that
\begin{equation}
\xi=\xi^\prime \quad \mathrm{and}\quad
\kappa_\perp^\prime=\kappa_\perp+(1-\xi)\, Q_\perp \quad
\mathrm{with}\quad Q_\perp=\left(
\begin{array}{c}Q\\0
\end{array}\right)\, .
\end{equation}
The invariant mass of the incoming (outgoing) $q\bar{q}$-system
can then be written as
\begin{equation}\label{eq:invmlc}
m_{q \bar{q}}^{(\prime)}=\sqrt{\frac{m_q^2+\kappa_\perp^{(\prime)
2}}{\xi (1-\xi)}}\, .
\end{equation}
With these relations the Jacobian for the variable transformation
$(\tilde{k}_q^{1\prime},\tilde{k}_q^{2\prime},\tilde{k}_q^{3\prime})
\rightarrow (\xi,\kappa_\perp^1,\kappa_\perp^2)$ becomes
\begin{equation}
\frac{\partial(\tilde{k}_q^{1\prime},\tilde{k}_q^{2\prime},\tilde{k}_q^{3\prime})}
{\partial(\xi,\kappa_\perp^1,\kappa_\perp^2)}=\frac{m_{q
\bar{q}}^{\prime}}{4 \xi (1-\xi)}\,
\end{equation}
and the integral for the electromagnetic form factor takes on the
form
\begin{equation}\label{eq:ffff}
F(Q^2) =\frac{1}{4 \pi }\int_0^1 \mathrm{d}\xi \int_{\mathbb{R}^2}
\mathrm{d}^2\!\kappa_\perp \frac{\sqrt{m_{q\bar{q}}\,
m'_{q\bar{q}}}}{4\xi(1-\xi)}\, \mathcal{M}\, u_{n 0}^\ast\,(\vert
\vec{\tilde{k}}_q^\prime\vert)\, u_{n 0}\,(\vert
\vec{\tilde{k}}_q\vert)
 \, .
\end{equation}
The argument of the wave functions is easily expressed in terms of
$\xi$ and $\kappa_\perp$ if one uses $\vec{\tilde{k}}_q^{(\prime)
2}=m_q^2+{m_{q \bar{q}}^{(\prime)\,2}}/{4}$ and
Eq.~(\ref{eq:invmlc}). By the change of variables the
spin-rotation factor $\mathcal{S}$ (cf. Eq.~(\ref{eq:spinrot}))
goes over into the Melosh-rotation factor
\begin{equation}\label{eq:melrot}
\mathcal{M}=\frac{m_{q \bar{q}}}{m_{q \bar{q}}^{\prime}}\left(
1+\frac{(1-\xi)\, (Q_\perp \cdot
\kappa_\perp)}{m_q^2+\kappa_\perp^2}\right)\, .
\end{equation}
$\mathcal{S}$ and $\mathcal{M}$ describe the effect of the quark
spin onto the electromagnetic form factor in point form and in
front form, respectively. Equations~(\ref{eq:ffff}) and
(\ref{eq:melrot}) are identical to the corresponding formulae in
Refs.~\cite{Chung:1988mu,Simula:2002vm}. This is a remarkable
result. Starting from two different forms of relativistic dynamics
and applying completely different procedures to identify the
electromagnetic meson form factor the outcome is the same. It
means that relativity is treated in an equivalent way and the
physical ingredients are alike. Since the infinite-momentum frame
we use is just a particular $q^+=0$ frame, Z-graph contributions
to the electromagnetic meson form factor are also suppressed in
our point-form approach. This is a welcome feature, because
Z-graphs can play a significant role in $q^+\neq 0$
frames~\cite{Simula:2002vm} and one should have control on them when
form-factor predictions are compared with experiment.

\subsection{\label{subsec:pfspectator}Comparison with the point-form spectator model}
Relativistic point-form quantum mechanics has also been applied in
Ref.~\cite{Wagenbrunn:2000es,Boffi:2001zb} to calculate
electroweak baryon form factors within a constituent quark model.
The strategy for the extraction of hadron form factors, however,
differs from the one in the present paper. The Bakamjian-Thomas
type framework which we apply to the full electron-hadron system
in order to derive the electromagnetic hadron current is only used
to obtain the bound-state wave function of the hadron. This wave
function is then plugged into an ansatz for the electromagnetic
hadron current. The ansatz is constrained by the requirements that
the hadron current should be conserved and should have the correct
properties under space-time translations and Lorentz
transformations. It is shown that these constraints can be
satisfied by a spectator current if not all of the photon momentum
is transferred to the struck quark. The momentum transfer to the
active quark $\tilde{Q}$ is uniquely determined by total
4-momentum conservation for the $\gamma^\ast H\rightarrow H $
subprocess and by the spectator conditions. An ambiguity in
defining such a spectator current, however, enters through a
normalization factor $\mathcal{N}$ which has to be introduced in
order to recover the hadron charge from the electric form factor
in the limit $Q^2\rightarrow 0$~\cite{Melde:2004qu}. Since both
quantities, $\tilde{Q}$ and $\mathcal{N}$ depend effectively on
all quark momenta and not only on those of the active ones, the model
current constructed in this way cannot be considered as a pure
one-body current~\cite{Melde:2007zz}. It has therefore been termed
\lq\lq point-form spectator model\rq\rq\ (PFSM) to distinguish it
from the usual impulse approximation.

\begin{figure}
\begin{center}
 \includegraphics[clip=5cm,width=10cm]{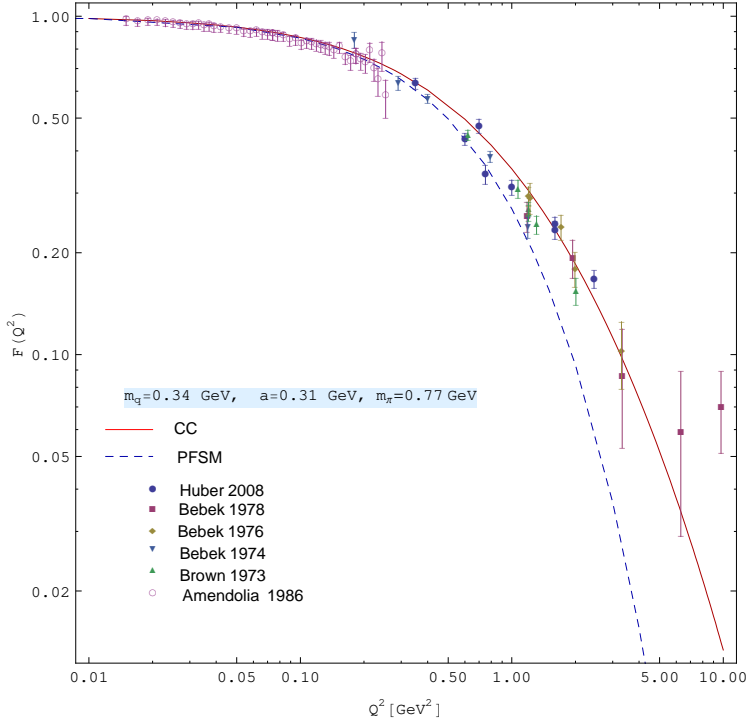}
 \caption{\label{fig:comp1}The electromagnetic pion form factor
 $F\left(Q^2\right)$ as evaluated by means of
 Eq.~(\ref{eq:formfactor}) (solid line) in comparison
 with the outcome of the point-form spectator model (dotted line).
 Data as in Figs.~\ref{fig:fflowQ2} and \ref{fig:ffhighQ2}.}
 \end{center}
 \end{figure}

Another characteristic feature of PFSM form factors  is that they are
determined not only by the hadron bound-state
wave function, but  also exhibit  an explicit dependence on the mass
of the bound state. Within the PFSM the eigenvalue spectrum of the
mass operator is thus directly connected with the electromagnetic
structure of its eigenstates. This makes it somewhat delicate to
compare our form-factor results for the simple harmonic-oscillator
confinement potential with corresponding PFSM predictions. Whereas
the wave function is solely determined by the oscillator parameter
$a$ (cf. Eq.~(\ref{eq:u00})), another free constant $V_0$ can be
added to the confinement potential to shift the spectrum. Unlike our
results, which do not depend on $V_0$, the PFSM predictions exhibit
a strong dependence on $V_0$. Taking $a=0.35$~GeV (as above) and
$V_0<0$, such that if the harmonic oscillator ground state were to coincide
with the physical mass of the pion, the fall-off of the form factor
would be unreasonably strong. With $V_0=0$, on the other hand, the
pion ground-state mass would be larger than $1$~GeV and the fall-off
of the form factor would be much too slow. We have therefore tried
to take another set of parameters for the harmonic-oscillator
confinement potential which is fixed through the
vector-meson spectrum~\cite{Krassnigg:2003gh}. With this set of
parameters, i.e. $a=0.312$~GeV, $V_0=-1.04$~GeV$^2$ and
$m_q=0.34$~GeV, the masses of the vector meson ground states and
first excited states are reasonably well reproduced. Applying them to
the $\pi$ meson and its excitations we observe that the first and
second radial excitations are about 10\% too high as compared with
experiment and the $\pi$ ground state has a mass of $0.77$~GeV.
These are reasonable values for a pure central confining potential
in view of the fact that spin-spin forces from an additional
hyperfine interaction can bring them close to the experimental
masses~\cite{Carlson:1983rw}. But for the purpose of a qualitative
comparison of our coupled-channel formalism with the PFSM we just
stay with the simple harmonic-oscillator $q$-$\bar{q}$ potential. If
it is parameterized as in Ref.~\cite{Krassnigg:2003gh} the
predictions for the electromagnetic $\pi$ form factor obtained by
means of Eq.~(\ref{eq:formfactor}) and with the PFSM become
comparable at small momentum transfers (cf.~Fig.\ref{fig:comp1}).
Above $Q^2\approx 1$~GeV$^2$, however, significant differences can
be observed. These differences resemble the situation for the
electromagnetic nucleon form factors. In the latter case the
stronger fall-off produced by the PFSM is a welcome feature which
brings the theoretical predictions from constituent quark models
close to experiments~\cite{Melde:2007zz}. For the usual front-form
spectator current in the $q^+=0$ frame agreement with experiment is
 achieved only by introducing electromagnetic form factors
for the constituent quarks~\cite{Simula:2001wx}. It remains to be
seen whether the situation for the electromagnetic $\pi$ form factor
could also change in favor of the PFSM if a more sophisticated
$q$-$\bar{q}$ potential is employed which reproduces the mass of the
$\pi$ meson and its lowest excitations sufficiently accurately.

% \noindent{\bf Effect of Wigner rotations, Log-Log Plot,
% Analytical
% expression for the PFMS FF in the IF-frame?}

\section{\label{sec:conclusion}Summary and outlook}
We have analyzed the electromagnetic structure of pseudoscalar
mesons by utilizing the point form of relativistic quantum
mechanics in connection with a velocity-state representation.  In this
 formulation the scattering of an electron by a confined
quark-antiquark pair becomes a two-channel problem
for a Bakamjian-Thomas type mass operator. With the second
channel, the $eq\bar{q}\gamma$ channel, the dynamics of the
exchanged photon is explicitly taken into account. Confinement is
treated via an instantaneous potential. The emission and
absorption of a photon by an electron or (anti)quark is described
by a vertex interaction which has the Lorentz structure of the
field theoretical vertex, but conserves the 4-velocity of the
whole $e q \bar{q}(\gamma)$ system. By construction this approach
is Poincar\'e invariant. After reduction of the eigenvalue problem
for the coupled-channel mass operator to a non-linear eigenvalue
equation for the $e q \bar{q}$ channel, the matrix elements of the
electromagnetic meson current can be read off from the
one-photon-exchange optical potential. As expected the optical
potential is a contraction of the point-like electron current with
a meson current times the covariant photon propagator. The
resulting meson current transforms as a 4-vector and
satisfies current conservation. The extracted meson form factor,
however, depends not only on $Q^2$, i.e. the 4-momentum
transfer squared, but also on Mandelstam $s$, i.e. the
total invariant mass squared of the electron-meson system. This
finding hints at a violation of cluster separability, since the
meson current is also influenced by the presence of the electron.
The violation of cluster separability is a consequence of the approximation
that the total velocity of the electron-meson system is conserved
throughout the electromagnetic scattering process. On the other hand, it
is only because of
this approximation that we have been able to formulate electron-meson
scattering as a simple eigenvalue problem for a Bakamjian-Thomas
type mass operator. A possible way to recover cluster separability
is the introduction of so called \lq\lq packing
operators\rq\rq~\cite{Sokolov:1977ym,Coester:1982vt,Keister:1991sb},
i.e. appropriate unitary transformations which restore cluster
separability. We have not made use of such packing operators, but
have rather exploited the observation that the $s$-dependence of
the meson form factor vanishes  rapidly with increasing $s$,
indicating that cluster-separability-violating effects become
negligible if the meson momentum is large enough. It has been
shown analytically that in the limit of infinitely large meson
momentum the current goes over into a product of the usual
point-like meson current times an integral with the integrand
depending only on internal variables (which are integrated and
summed over) and on the momentum transfer $Q$. The limit of the
current is a one-body current. The physical meaning of letting the
meson momentum go to infinity is that the $\gamma^\ast M
\rightarrow M$ subprocess is considered in the infinite momentum
frame of the meson. We have thus succeeded in deriving a conserved
electromagnetic current for a composite pseudoscalar meson which
acquires the correct covariance properties under Poincar\'e
transformations if the meson momentum is sufficiently large. It
should be emphasized that this is no restriction on the
(space-like) momentum transfer $Q$. Finally we have been able to
prove that our analytical formula for the form factor is
equivalent to the usual front-form expression that results from a
spectator current in the $q^+=0$ frame.

 Apart from
confirming that the front- and point-form of relativistic dynamics
 lead to equivalent results for the electromagnetic form
factor of a pseudoscalar meson, it might seem that there is
 nothing new in these results. What makes our approach
interesting, however, is that it can immediately be generalized
in various directions. It is quite obvious how to proceed
for (space-like) form factors of arbitrary few-body systems which
are bound by an instantaneous potential. Even more, one could
account for dynamical particle-exchange interactions by adding
additional channels. In this case the current of the bound
few-body system would not be a simple one-body current, but would
also contain many-body contributions. In general, the construction
of such many-body currents is a highly nontrivial
task~\cite{Gross:1987bu}. The advantage of our approach is that
the current is uniquely determined by the interaction dynamics
which is responsible for the binding and  can be read
off directly from the one-photon exchange optical potential in the
electron-cluster channel. The only problem with this
procedure is connected with cluster separability. For the simple
example treated in this paper this problem has been overcome by
going to the infinite-momentum frame of the cluster. It remains
to be seen whether the same strategy also works in the general
case, or whether one has to apply packing operators to ensure
cluster separability right from the beginning.

 In principle, one
could also think of applying our coupled-channel formalism to the
calculation of time-like hadron form factors. One has to study the
process $e^- e^+ \rightarrow \gamma^\ast \rightarrow H \bar{H}$.
In this case, however, the situation becomes much more
complicated. It is not enough to know the bound-state dynamics of
the hadron $H$. It is also necessary to specify the (strong)
interaction mechanism that produces those hadronic constituents
which do not directly couple to the photon. In the time-like case
the argument of the form factor is Mandelstam $s$. It is therefore
not possible to go into the infinite momentum frame in order to
get rid of problems with cluster separability, which most likely
will show up as additional angular dependence of the form factors.
A possible way out could again be the application of packing
operators.

To summarize, we have presented a relativistic formalism which
makes it possible to derive the electromagnetic current and form
factors of a bound few-body system consistent with the binding
forces. In this paper the formalism has been worked out in detail
for a quark-antiquark pair with the quantum numbers of a pion
which is confined via an instantaneous potential. The formalism,
however, is much more general and applications to other systems
with more complicated bound-state dynamics will be the subject of
further investigations.

\begin{acknowledgments}
We would like to thank F. Coester, T. Melde and W. Plessas for
helpful discussions. E.B. acknowledges the support of the \lq\lq
Fond zur F\"orderung der wissenschaftlichen Forschung in
\"Osterreich\rq\rq (FWF DK W1203-N08).
\end{acknowledgments}

%
%\bibliography{pointformff}% Produces the bibliography via BibTeX.

\end{document}